\documentclass{emulateapj}
\usepackage[]{graphicx,enumerate}

\shorttitle{The RSR Observations of CO in Local ULIRGs}
\shortauthors{Chung et al.}
\begin{document}

\title{The Redshift Search Receiver Observations of {$^{12}$}CO~$J=1\rightarrow0$
       \\in 29 Ultraluminous Infrared Galaxies}

\author{Aeree Chung\altaffilmark{1,2}, Gopal Narayanan\altaffilmark{2},
        Min S. Yun\altaffilmark{2}, Mark Heyer\altaffilmark{2} 
        and Neal R. Erickson\altaffilmark{2}}
\altaffiltext{1}{Jansky Fellow, National Radio Astronomy Observatory; 
                 email: achung@aoc.nrao.edu}
\altaffiltext{2}{Department of Astronomy, University of Massachusetts, 
                 710 North Pleasant Street, Amherst, MA 01003, USA;
                 emails: gopal@astro.umass.edu, myun@astro.umass.edu, 
                 heyer@astro.umass.edu, neal@astro.umass.edu}

\begin{abstract}
We present $^{12}${\sc CO} ${J=1\rightarrow 0}$ observations of ultraluminous 
infrared galaxies (ULIRGs) obtained using the Redshift Search Receiver (RSR)
on the 14-m telescope of the Five College Radio Astronomy Observatory. The RSR
is a novel, dual-beam, dual-polarization receiver equipped with an ultra-wideband
spectrometer backend that is being built as a facility receiver for the Large
Millimeter Telescope. Our sample consists of 29 ULIRGs in the redshift range
of 0.04-0.11, including 10 objects with no prior $^{12}${\sc CO} measurements.  
We have detected 27 systems (a detection rate of 93\%), including 9 ULIRGs 
that are detected in {\sc CO} for the first time. Our study has increased the 
number of local ULIRGs with {\sc CO} measurements by $\sim15\%$. The {\sc CO} 
line luminosity $L_{\rm CO}^\prime$, correlates well with far-infrared luminosity 
$L_{FIR}$, following the general trend of other local ULIRGs. However, compared 
to previous surveys we probe deeper into the low {\sc CO} luminosity end of ULIRG 
population as a single study by including a number of {\sc CO} faint objects in 
the sample. As a result, we find 1) a smoother transition between the ULIRG 
population and local QSOs in $L_{FIR}-L_{\rm CO}^\prime$ (``star formation 
efficiency'') space, and 2) a broader range of $L_{FIR}/L_{\rm CO}^\prime$ 
flux ratio ($\sim$60--$10^3~L_\odot/$[K~km~s$^{-1}~$pc$^2$]) than previously 
reported. In our new survey, we also have found a small number of ULIRGs with 
extreme $L_{FIR}/L_{\rm CO}^\prime$ which had been known to be rare. The mid-$IR$ 
color and radio-excess of 56 local ULIRGs as a function of $FIR$-to-{\sc CO} flux
ratio is examined and compared with those of spirals/starburst galaxies and 
low-$z$ QSOs. In this paper, using a large sample of local ULIRGs we explore
the origin of their current power source and potential evolution to QSOs.
\end{abstract}

\section{Introduction}
\label{sec-int}
An important discovery from the $IRAS$ all-sky survey  is a large population of 
infrared luminous ($L_{IR}\gtrsim 10^{11}~L_\odot$) galaxies that are emitting 
the bulk of their luminosity in the infrared \citep{sm96}. The most extreme
objects, so-called ``ultraluminous infrared galaxies (ULIRGs)'' with 
$L_{IR}\gtrsim 10^{12}~L_\odot$, are the most luminous objects in the local 
Universe \citep{solomon97}. ULIRGs are rare at low redshift ($z\lesssim0.3$), 
but their co-moving $IR$ energy density grows rapidly with increasing redshift 
\citep{lefloch05}, and galaxies with ULIRG-like luminosity are quite numerous
at $z\gtrsim1$ \citep[][and references therein]{farrah03}. ULIRGs therefore
represent an important population for understanding the cosmic star formation
history and the cosmic energy budget. 

Most ULIRGs appear to be advanced mergers \citep{sanders88a,farrah01} while 
$\sim$15\% of $IR$ luminous systems show characteristics of quasi-stellar objects 
(QSOs) in their optical spectra \citep[][]{sm96}. Whether the origin of their 
large $IR$ luminosity is a merger-driven starburst or AGN activity (or both)
is still not well established. Their eventual fate is also controversial. In the
multi-wavelength study of a flux-limited sample, \citet[][b]{sanders88a} have
suggested that ULIRGs hosting an AGN are transition objects that eventually evolve
into optically selected QSOs as the AGN light dominates the decaying
starburst light. 

A common trait found among the ULIRG population is the large molecular gas
content ($M_{H_2}\ga 10^{10} M_\odot$), inferred from their observed CO
luminosity \citep{sanders91,young95,solomon97}, capable of fueling both the
bursts of star formation and the AGN activities seen in these systems. 
Indeed, the correlation between the far-infrared luminosity $L_{FIR}$ and the 
{\sc CO} line luminosity $L_{\rm CO}^\prime$ of ULIRGs forms a continuous
track with those of QSOs in the Local Universe
\citep[][and references therein]{riechers06}, which is supportive of the 
proposed ULIRG-QSO evolution scenario. QSOs with $L_{IR}>10^{12}~L_\odot$
however are rare in the Local Universe, implying that ULIRGs are not simply
a stage of transition to QSOs \citep{farrah01,yun04} or that the transition
may occur quite rapidly \citep{yun04}.

Considering the rich history of multi-wavelength investigations of the ULIRG
phenomenon and their importance as the local analogs to the luminous dusty
galaxies found in the early universe \citep[so-called ``submillimeter 
galaxies (SMGs)''; see review by][]{blain02}, it is surprising that the total
number of published {\sc CO} measurements for ULIRGs is limited to only 
$\sim$50 or so, primarily from the surveys conducted by \citet{sanders91} 
and \citet{solomon97}\footnote{More recent ULIRG surveys of the higher
$J$ CO transitions \citep{yao03,narayan05} have targeted mostly the same
sample.}, and by \citet{mirabel90} for southern objects. There have
been also some {\sc CO} studies of individual ULIRGs in the Local Universe
\citep[$z\lesssim0.3$, e.g.][]{combes06}, although these do not contribute
to the total number statistics by large. 

This limited information on {\sc CO} emission and molecular gas content reflects
both the rarity of ULIRGs in the local volume and the challenge of detecting 
CO emission from galaxies even at these modest distances. It raises a real
concern that the existing {\sc CO} data may be too small to characterize the 
molecular gas properties of the ULIRG population with sufficient statistics.
The existing {\sc CO} data may also reflect strong biases in the sample selection 
and/or the survey strategies adopted by these earlier studies. In comparison,
the total number of {\sc CO} measurements available for the $z>1$ SMGs is $>$30
systems \citep[see review by][]{svb05}, which is generally considered too few
to draw any broad conclusions about the molecular gas content and properties
of the SMG population.

In order to address these issues, we have conducted a new {\sc CO} survey of 
29 ULIRGs in the redshift range $z\approx0.04$-0.11 using the sensitive, 
ultra-wideband spectrometer system called the Redshift Search Receiver (RSR)
on the Five College Radio Astronomy Observatory (FCRAO) 14-m telescope. 
The broad spectral coverage of the RSR enables a {\sc CO} line detection even in 
cases where the reported optical redshifts might have been significantly in
error.  Our sample is comparable in size to the largest previous {\sc
CO} surveys of ULIRG population \citep[e.g.][]{solomon97}, but probes deeper 
into the {\sc CO} luminosity by including more {\sc CO} faint objects in 
the sample as a single study. Our sensitivity allowed us to 
detect a number of sources with the comparable line intensity as the faintest
ULIRGs in {\sc CO} that have been known to date (e.g. IRAS~08572$+$3915 by 
Solomon et al. 1997 or IRAS~10173$+$0828 by Sanders et al. 1991).

%\placetable{tbl-sample}
\begin{table*}
\begin{center}
\caption{\sc General Properties of the Sample of 29 ULIRGs\tablenotemark{a}\label{tbl-sample}}
\begin{tabular}[b]{llrrcccrcll}
\hline\hline
\multicolumn{1}{l}{IRAS name} &
\multicolumn{1}{c}{Other Name} &
\multicolumn{1}{c}{~~$\alpha_{2000}$} &
\multicolumn{1}{c}{~~$\delta_{2000}$} &
\multicolumn{1}{c}{$cz$} &
\multicolumn{1}{c}{$m_B$} &
\multicolumn{1}{c}{$D_{opt}$} &
\multicolumn{1}{c}{$t_{\rm int}$} &
\multicolumn{1}{c}{rms} &
\multicolumn{1}{l}{Morphology} & 
\multicolumn{1}{l}{Reference\tablenotemark{b}}\\
\multicolumn{1}{c}{} &
\multicolumn{1}{c}{} &
\multicolumn{1}{c}{($hh mm ss.s$)} &
\multicolumn{1}{c}{~~~($^\circ~~'~~''$)} &
\multicolumn{1}{c}{(km s$^{-1}$)} &
\multicolumn{1}{c}{(mag)} &
\multicolumn{1}{c}{($''$)} &
\multicolumn{1}{c}{(hrs)} &
\multicolumn{1}{c}{(mK)} &
\multicolumn{1}{c}{} &
\multicolumn{1}{c}{} \\
\hline
00057$+$4021 &PGC 000626  & 00 08 20.5 & 40 37 57 &13389 &16.78 &0.46 &  2.9~ & 0.50 &Sy2 & S97\\
05083$+$7936 &VII Zw 031  & 05 16 46.4 & 79 40 13 &16090 &15.80 &...  &  1.9~ & 0.65 &H{\sc ii} & S97 \\
05189$-$2524 &            & 05 21 01.4 &-25 21 46 &12760 &15.40 &0.46 &  3.6~ & 0.65 &Pec Sy2 & M90, S91 [KS98]\\
08572$+$3915 &            & 09 00 25.0 & 39 03 54 &17493 &14.92 &...  &  8.2~ & 0.31 &LINER; Sy2& S97, GS99 [KS98, M01]\\
09111$-$1007 &LEDA 153577 & 09 13 38.8 &-10 19 20 &16231 &16.11 &0.70 &  2.5~ & 0.53 &H{\sc ii} Sy2& M90 [M01]\\
10035$+$4852 &PGC 029385  & 10 06 46.2 & 48 37 46 &19430 &15.70 &0.80 &  1.5~ & 0.57 &Sbrst& S97 [KS98, M01]\\
10173$+$0828 &PGC 030202  & 10 20 00.2 & 08 13 34 &14716 &16.90 &0.44 &  3.7~ & 0.38 &Megamaser& S91\\
10190$+$1322 &LEDA 090120 & 10 21 42.5 & 13 06 54 &22955 &17.51 &0.30 &  5.7~ & 0.31 &H{\sc ii}& S97 [KS98, M01]\\
10494$+$4424 &LEDA 090133 & 10 52 23.2 & 44 08 48 &27599 &17.40 &0.34 & 10.8~ & 0.25 &LINER& S97 [KS98, M01]\\
10565$+$2448 &PGC 033083  & 10 59 18.1 & 24 32 34 &12921 &15.70 &0.40 &  1.4~ & 0.60 &LINER; H{\sc ii}& S97\\
11095$-$0238 &            & 11 12 03.4 &-02 54 22 &31968 &17.45 &0.17 &  6.3~ & 0.38 &Merger LINER& [KS98, M01]\\
12112$+$0305 &            & 12 13 46.0 & 02 48 38 &21980 &16.90 &0.50 &  1.2~ & 0.63 &LINER; H{\sc ii}& S91 [KS98, M01]\\
12540$+$5708 &Mrk 231     & 12 56 14.2 & 56 52 25 &12642 &14.41 &1.30 &  2.5~ & 0.55 &SA(rs)c? pec Sy1& S91, S97 [KS98]\\
13539$+$2920 &LEDA 84081  & 13 56 10.0 & 29 05 35 &32513 &16.90 &0.40 &  7.0~ & 0.27 &H{\sc ii}:& [KS98]\\
14348$-$1447 &PGC 052270  & 14 37 38.3 &-15 00 22 &24802 &16.58 &0.60 &  3.8~ & 0.82 &Merger: LINER& M90, S91 [KS98, M01]\\
14394$+$5332 &LEDA 084264 & 14 41 04.4 & 53 20 09 &31341 &17.20 &0.21 &  4.2~ & 0.36 &Sy2& [KS98, M01]\\
15130$-$1958 &LEDA 090203 & 15 15 55.2 &-20 09 17 &32792 &17.10 &...  &  2.0~ & 0.73 &Sy2& [KS98]\\
15250$+$3609 &PGC 055114  & 15 26 59.4 & 35 58 38 &16535 &16.20 &0.44 &  5.2~ & 0.38 &Ring; LINER& [M01]\\
15462$-$0450 &LEDA 090222 & 15 48 56.8 &-04 59 34 &29917 &16.40 &...  &  3.4~ & 0.46 &Sy1& [KS98, M01]\\
16487$+$5447 &LEDA 090261 & 16 49 47.0 & 54 42 34 &31106 &16.50 &...  &  6.1~ & 0.31 &Sbrst LINER& GS99 [KS98, M01]\\
17028$+$5817 &LEDA 090270 & 17 03 41.9 & 58 13 44 &31805 &17.12 &...  &  3.8~ & 0.50 &E;LINER;Sbrst H{\sc ii}& [KS98, M01]\\
17132$+$5313 &PGC 059896  & 17 14 20.0 & 53 10 30 &15270 &...   &...  &  2.5~ & 0.46 &H{\sc ii}& S91, GS99\\
17208$-$0014 &PGC 060189  & 17 23 21.9 &-00 17 01 &12834 &15.10 &0.40 &  2.5~ & 0.67 &Sbrst H{\sc ii}& M90, S97\\
18470$+$3233 &            & 18 48 53.8 & 32 37 28 &23517 &16.00 &0.10 &  4.0~ & 0.42 &Pair H{\sc ii}& [M01]\\
19297$-$0406 &            & 19 32 21.2 &-03 59 56 &25701 &16.00 &0.22 &  5.7~ & 0.37 &Merger H{\sc ii}:& S97\\
20414$-$1651 &            & 20 44 18.2 &-16 40 16 &26107 &17.10 &...  &  7.8~ & 0.34 &LINER; H{\sc ii} Sbrst& M90 [KS98, M01]\\
22491$-$1808 &PGC 069877  & 22 51 49.0 &-17 52 23 &23312 &16.15 &0.30 &  5.1~ & 0.32 &Merger; H{\sc ii}& M90, S91 [KS98, M01]\\
23327$+$2913 &            & 23 35 11.9 & 29 30 00 &32078 &16.80 &...  &  4.3~ & 0.36 &LINER& [KS98, M01]\\
23365$+$3604 &LEDA 090429 & 23 39 01.3 & 36 21 09 &19331 &16.30 &0.46 &  4.3~ & 0.45 &SBa? pec LINER& S97 [M01]\\
\hline
\end{tabular}
\end{center}
\footnotetext{}{$^{\rm a}$These data have been compiled from the NASA/IPAC Extragalactic Database (NED) 
which is operated by the Jet Propulsion Laboratory, California Institute of Technology, under 
contract with the National Aeronautics and Space Administration.}\\
\footnotetext{}{$^{\rm b}$References for previous {\sc CO} measurements: 
M90-\citet{mirabel90}; S91-\citet{sanders91}; S97-\citet{solomon97}; 
GS99-\citet{gs99} [References for the optical studies:
KS98-\citet{ks98}'s 1Jy sample; M01-\citet{murphy01}'s 2~Jy sample]}
\end{table*}

This paper is organized in the following order. In \S\ref{sec-sam} we describe
our sample selection criteria and the properties of the sample. In \S\ref{sec-obs}
we describe the observations and data reduction. In \S\ref{sec-res} we present
the results and discuss the data quality. In \S\ref{sec-dis} the origin of $IR$
emission and the evolution of the local ULIRGs are discussed. Finally in
\S\ref{sec-sum} we summarize our findings. We assume a standard $\Lambda$CDM
cosmology with $H_0=$71~km~s$^{-1}$~Mpc$^{-1}$, $\Omega_M=$0.27, and
$\Omega_\Lambda=$0.73 \citep{cosmos03} throughout the paper.

\begin{figure*}
\plotone{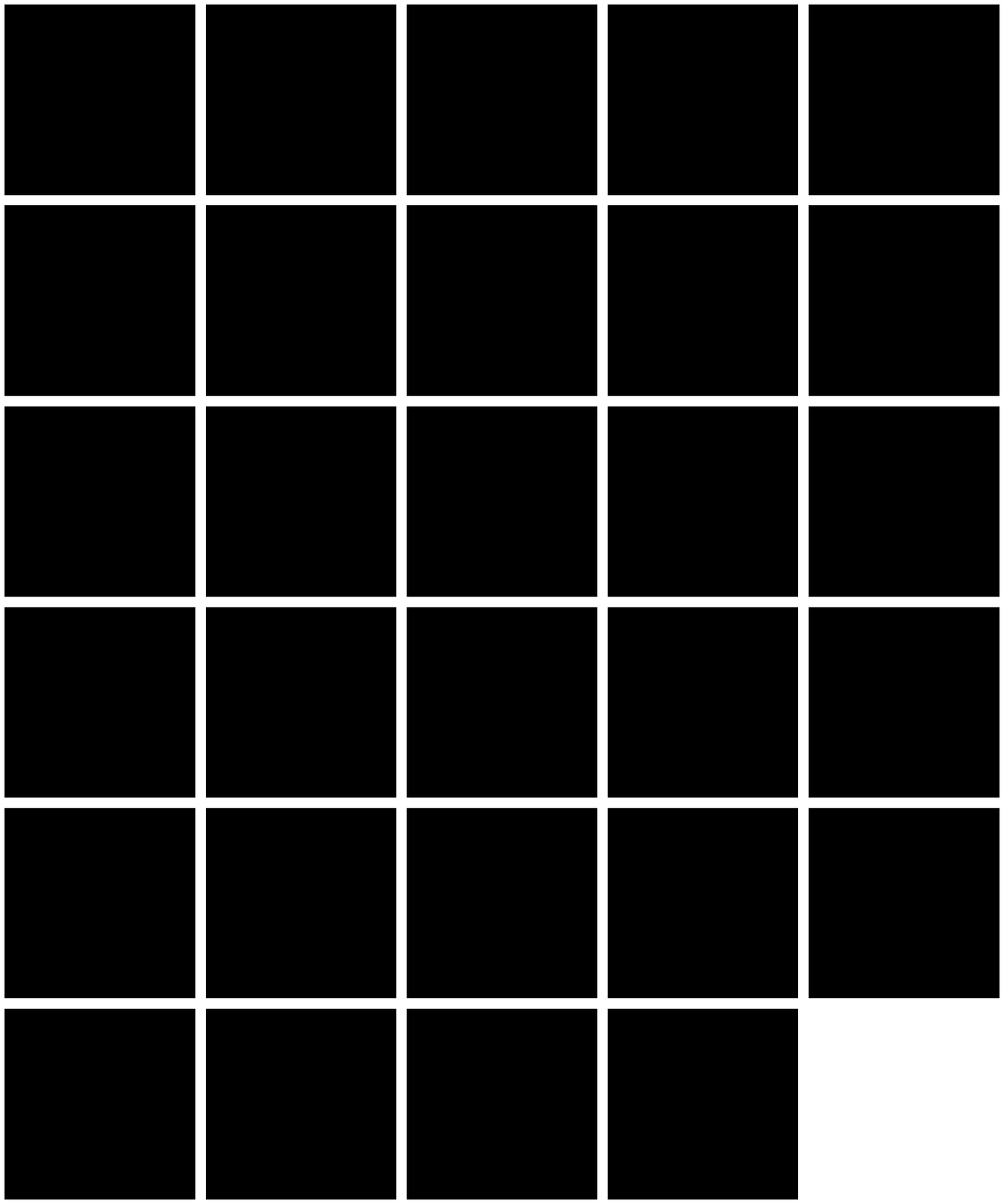}
\caption{Optical images from the 2nd Digitized Sky Survey (DSS2) of the sample
of 29 local ($z=0.043$-0.11) ULIRGs. Each field (1.5$'\times1.5'$) is centered on 
($\alpha,\delta$)$_{2000}$ as shown in Table~\ref{tbl-sample}. The $IRAS$ catalog number
is indicated on the top of each panel. A circle of $50''$ ($\approx$beamsize of 14m at 3mm)
is shown in solid line. \label{fig-sample}}
\end{figure*}

\placefigure{fig-sample}

\section{Sample}
\label{sec-sam}
Our sample is constructed primarily from the 1~Jy ULIRG\footnote{The conventional
definition of an ultraluminous infrared galaxy is a galaxy with rest-frame 1-1000
$\micron$ luminosity of $L_{IR}\geq10^{12}~L_\odot$ \citep[see][]{sm96}.} sample,
which consists of 118 sources identified from the IRAS Faint Source Catalog
\citep{moshir90} with $S_{60\mu}>1$ Jy, $\delta>-40^\circ$, and $|b|>30^\circ$
\citep{ks98}. All 18 ULIRGs in the 1~Jy Sample whose $^{12}${\sc CO} $J=1\rightarrow0$
line falls within the frequency range of $\nu_{\rm obs}\approx104$-111~GHz
($z=$0.043-0.11, see \S\ref{sec-obs}) were selected as the primary
sample. In addition, we have selected four targets from the 2~Jy
sample of \citet{murphy01} with $S_{60\mu}>1.94$ Jy, $\delta>-35^\circ$, 
and $|b|>5^\circ$. Lastly, in order to provide comparison with our
observations (\S~\ref{sec-quality}), 20 ULIRGs with previous {\sc
CO} measurements were added from \citet{mirabel90}, \citet{sanders91},
\citet{solomon97}, and \citet{gs99}. Those ULIRGs were all previously
detected except for IRAS~20414$-$1651 \citep{mirabel90}.
These include 13 sources which overlap with the sample from
\citet{ks98} or \citet{murphy01}, or both, yielding the total number
of our sample 29. General properties of the 29 ULIRGs
are summarized in Table~\ref{tbl-sample}.

\begin{figure*}
\plotone{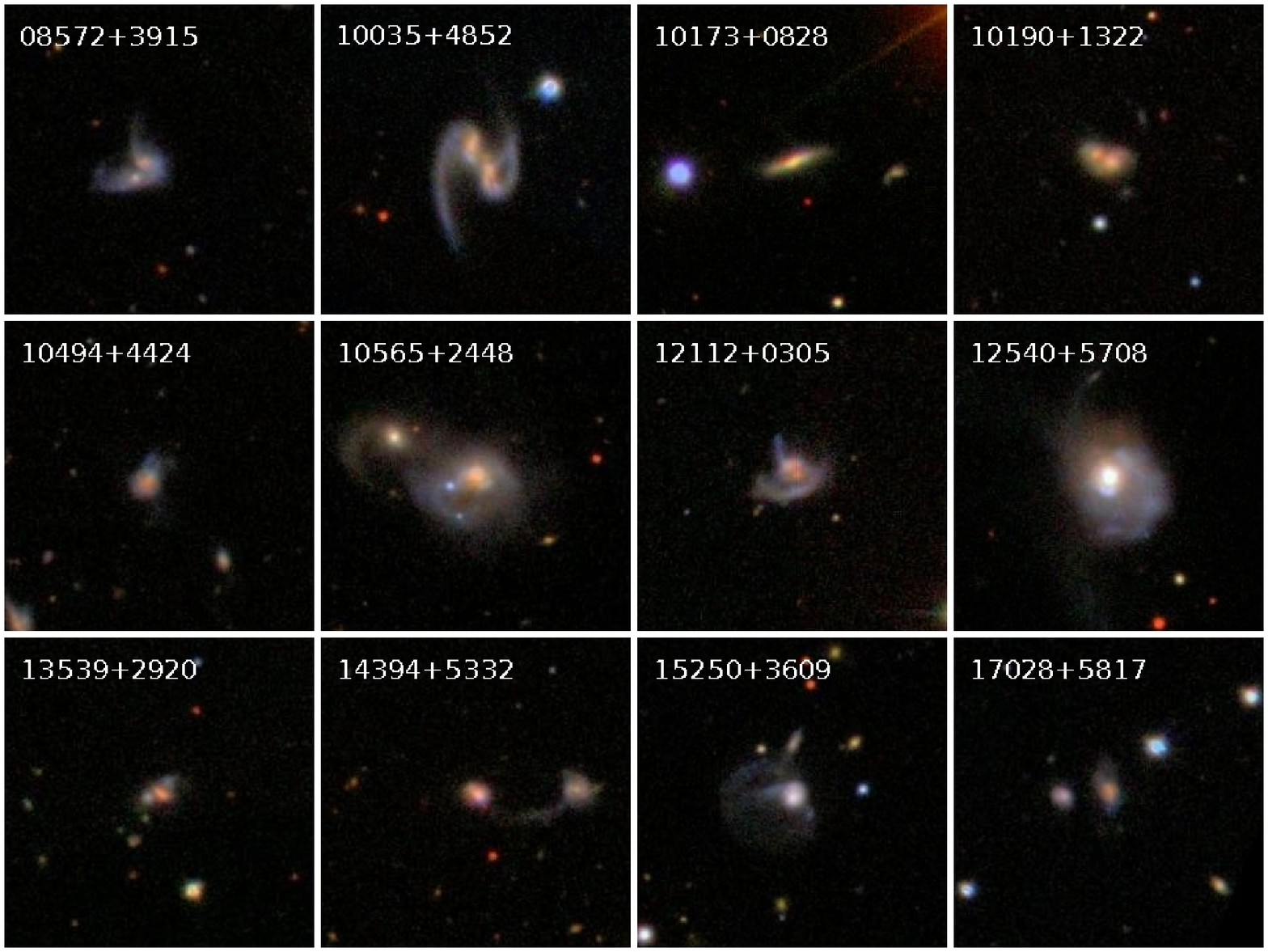}
\caption{The optical color images from the Sloan Digital Sky Survey
(SDSS, {\tt http://www.sdss.org/}) of 12 ULIRGs. The size of each field
is the same as Figure \ref{fig-sample}. \label{fig-sdss}}
\end{figure*}

\placefigure{fig-sdss}

The optical Digitized Sky Survey plate images of all 29 ULIRGs and 
the color images of 12 available ULIRGs from the Sloan Digital
Sky Survey data archive are presented in Figure~\ref{fig-sample} and 
\ref{fig-sdss}. As shown in the optical images, morphological
peculiarities are quite common in ULIRGs. 
Many galaxies show faint stellar tails (e.g. IRAS~12112$+$0305), 
rings (IRAS~15250$+$3609), or extended/asymmetric low surface brightness disks
(e.g. IRAS~10565$+$2448). More than 75\% of the sample are classified as
{\sc Hii}, LINER or a starburst systems. Three objects
IRAS~05189$-$2524, IRAS~08572$+$3915, and IRAS~12540$+$5708 (Mrk~231) have 
previously been identified as {\lq\lq}$warm${\rq\rq} ULIRGs with 
$f_{25}/f_{60}>0.2$ by \citet{sanders88b}, and another three 
(IRAS~14394$+$5332, IRAS~15130$-$1958, and IRAS~15250$+$3609) have warm $IR$ color 
($0.18\lesssim f_{25}/f_{60} \lesssim 0.2$), which is indicative of active galactic
nuclei \citep[AGN;][]{degrijp85}. In addition, IRAS~00057$+$4021 and
IRAS~15462$-$0450 are classified as Seyfert type 2 and 1, respectively, 
and likely harbor an AGN.  

Our study is one of the most systemic and deepest {\sc CO} surveys of 
a fairly large sample of ULIRGs to date, including 9 objects with no 
prior {\sc CO} measurments and one which was observed but not detected.

\section{Observations and Data Reduction}
\label{sec-obs}
\subsection{Observations with the RSR on the FCRAO 14-m Telescope}
The observations presented here were conducted using the Five College Radio Astronomy
Observatory (FCRAO) 14-m telescope between March and May in 2007 and in 2008 May. The
FCRAO 14-m telescope is a radome-enclosed single-dish millimeter telescope, located
on Prescott peninsula within the Quabbin Reservoir, Massachusetts. The
spectra were taken using the Redshift Search Receiver (RSR). The RSR is 
a sensitive, ultra-wideband spectrometer that is being built at the University of 
Massachusetts as one of the facility instruments for the 50-m diameter
Large Millimeter Telescope (LMT).

The RSR consists of a frontend receiver which uses a novel construction for a
millimeter wavelength system and a set of wide-band analog auto-correlation
spectrometers. There are four receivers each covering 74-111~GHz instantaneously
in a dual-beam, dual-polarized system. The input includes a novel electrical beam
switch which operates at 1 kilohertz (kHz) to overcome the 1/f noise originating 
within the frontend amplifiers as well as atmospheric noise to ensure excellent
baseline stability. The frontend uses MMIC (monolithic microwave integrated circuit) 
amplifiers and two very wideband mixers to convert each receiver band to two
intermediate frequency (IF) channels. After further conversion, the IF
signal passes into a spectrometer based on analog auto-correlation. 
Sets of tapped delay lines sample and multiply the signal with progressive
delays to generate a spectrum with 31~MHz resolution. Six spectrometers,
each covering 6.5~GHz bandwidth, are used with each pixel. The entire 36 GHz of
bandwidth of each pixel is handled by six boards, each having 256 lags (in all
1536 lags per pixel). Occasionally, a few of the lags develop problems,
which are flagged and blanked out before converting to the frequency domain.
However, less than 1\% 
of all lags had such problems, and most of these faulty lags have since then been
fixed in the spectrometer when the receiver was taken off the telescope
More details of the RSR and the LMT can be found in \citet{erickson07}.

During the observing season in 2007, all four frontend pixels (receivers) were 
available while only 4 of the 24 backend spectrometers had been fabricated.
This yielded a set of 4-frontend pixels $\times$ 1-backend spectrometer with a
bandwidth range of 104-111~GHz (1st LO could be tuned), and this spectral
coverage ultimately limited the redshift range in our sample selection.
In 2008, 12 spectrometers became available, and we had a set
of 4-pixels $\times$ 3-spectrometers with an instantaneous frequency 
coverage of 92-111~GHz.  Therefore some of our new spectra cover a much
larger spectral (redshift) range, but we focus on the analysis of 
only the $^{12}$CO $J=1\rightarrow0$ transition in this paper.  

We integrated between 1 and 11 hours (Table~\ref{tbl-sample}) on each
source depending on the weather conditions and the peak of the {\sc
CO} intensity. The system temperature during the run varied from 165 
to 380~K with a typical value of $T_{sys}\sim200~$K in both years. 
For calibration, normally we obtained one 5-min scan on the reference
sky every three consecutive 5-min scans on the target. When the weather was
bad ($T_{sys}\gtrsim400$~K), we did calibration every other scan
although most scans obtained under those weather conditions were 
thrown away in the final co-addition.
As a result, we achieved a
typical rms sensitivity of $\sigma \lesssim0.5$-0.6~mK in $T_A^*$ 
(corrected antenna temperature for the atmosphere absorption and
spillover losses). We checked the pointing and the focus using
planets every 2 to 3 hours. The antenna gain was monitored by measuring the 
intensity of planets.  An average gain factor of G$=45~$Jy~K$^{-1}$ is
adopted (in $T_A^*$ scale).

%\placefigure{fig-rsr-a}
\begin{center}
\begin{figure*}
\includegraphics[bb=50 245 560 775]{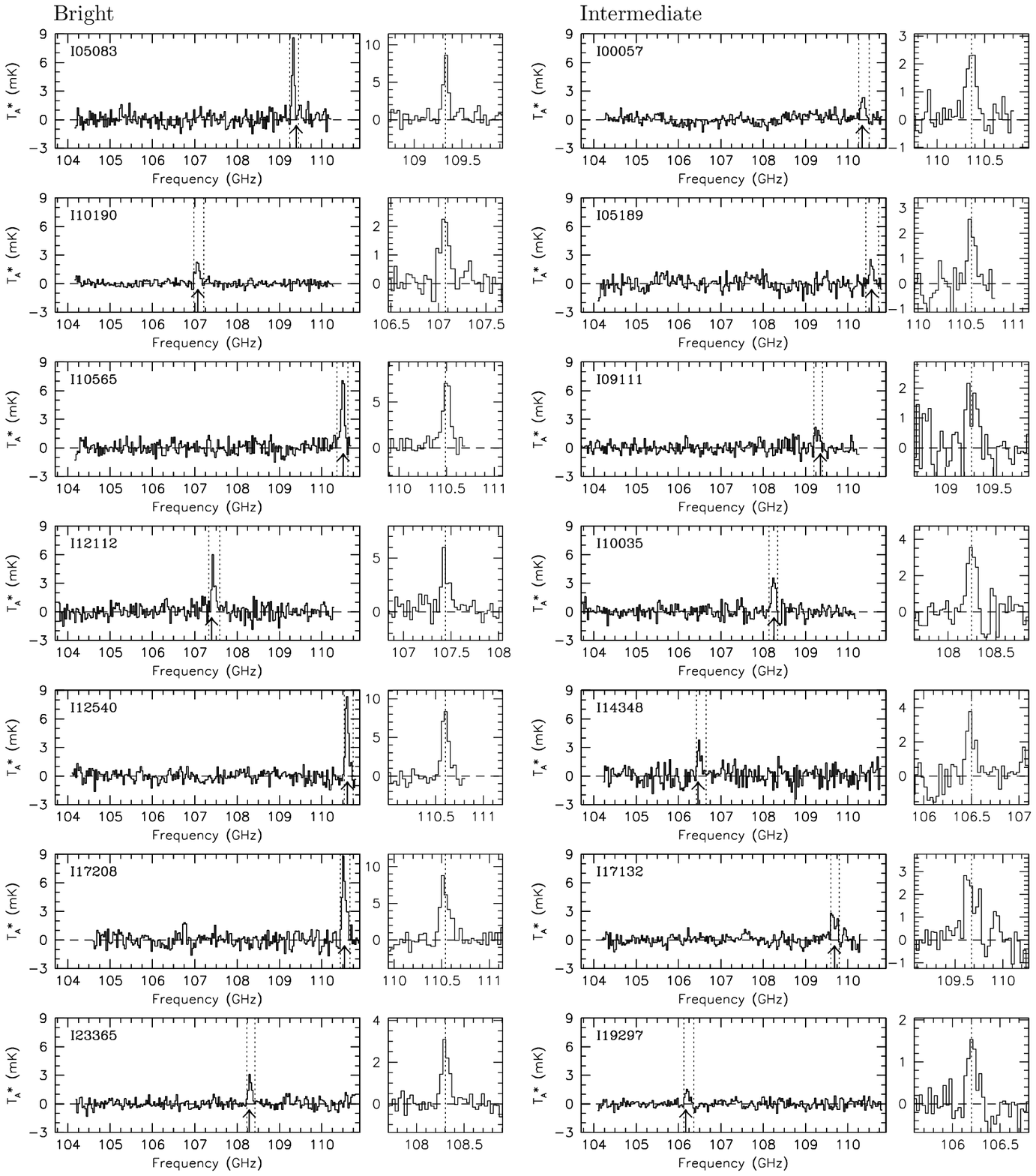}
\caption[]{(a) The RSR spectra of 14 ULIRGs with a high (left column) 
to intermediate (right column) CO strength. The quantitative descriptions
for the classiﬁcation are given in \S\ref{sec-res}.2. On the left side of 
each ob ject, the full abdn spectrum is shown except for a few noisy channels
at the edge. On the right-hand side, each spectrum is zoomed-in around the {\sc CO}
emission. The full-band spectra with the frequency range of 103.7-110.9 GHz are 
shown in the same $T^∗_A$ (antenna temperature corrected for the atmosphere 
absorption and spillover losses) scale for all galaxies. The dotted lines in 
each full-band spectrum represent the frequency range used to measure the CO 
intensity and derive the CO line width. The upper arrow indicates the estimated
CO frequency from the optical redshift. The zoom-in spectra of 1.2 GHz 
($\sim$3100 km~s$^{-1}$) width centered on the CO frequency are scaled down 
with the peak of the CO emission. The dotted line in the 
zoom-in spectrum represents the centroid CO frequency. \label{fig-rsr-a}}
\end{figure*}
\end{center}

%\placefigure{fig-rsr-b}
\addtocounter{figure}{-1}
\begin{figure*}
\includegraphics[bb=50 390 560 775]{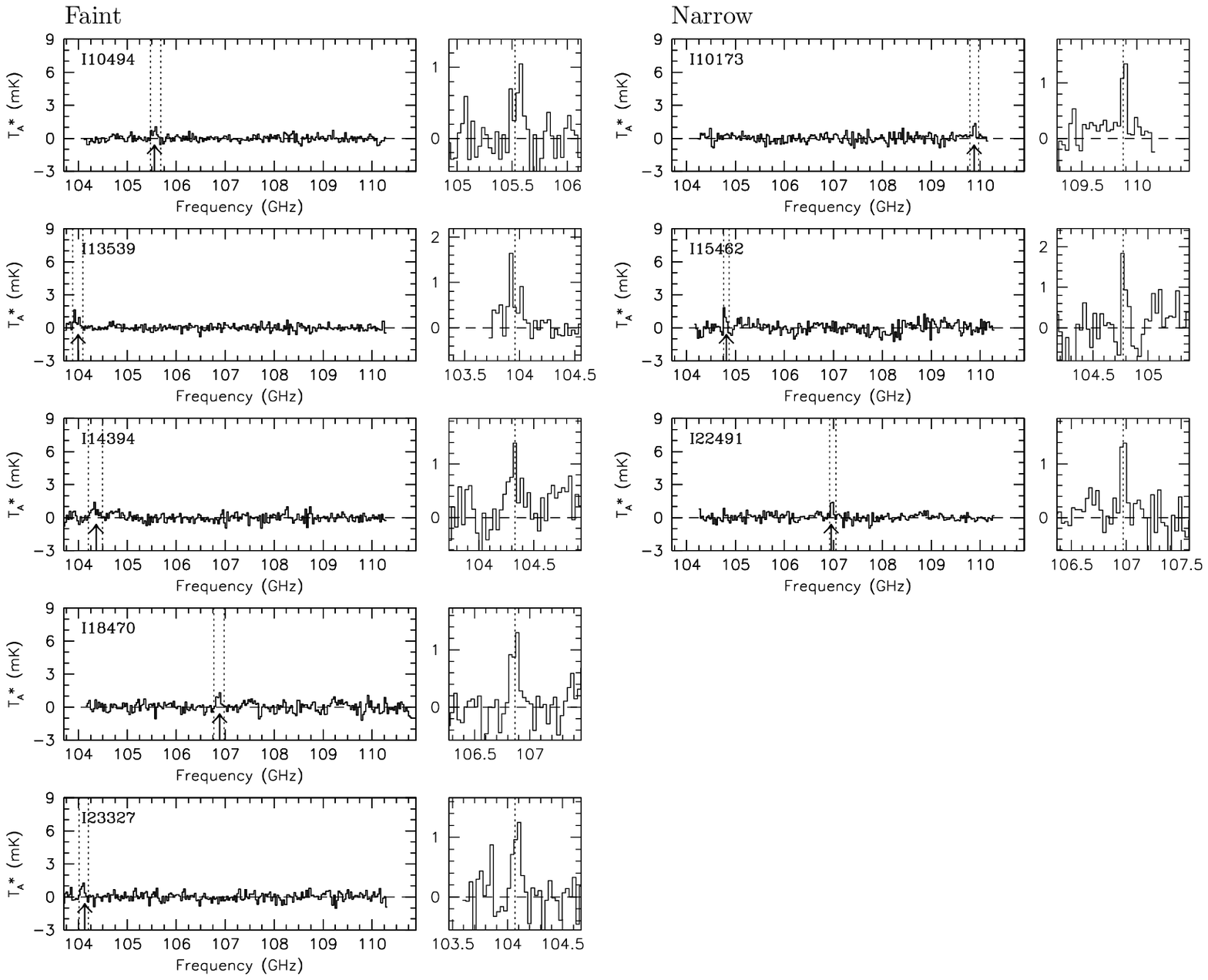}
\caption[]{(b) The RSR spectra of 5 ULIRGs that are faint in {\sc CO} (left 
column) and 3 ULIRGs that are faint and narrow (right column). 
See the caption of Figure~\ref{fig-rsr-a}-(a) for more details. 
\label{fig-rsr-b}}
\end{figure*}

\subsection{Data Reduction Using SPAPY}
The RSR data have been reduced using the {\tt SPAPY} ({\tt SPA} $+$ {\tt PYthon}).
{\tt SPAPY} is a data reduction software that has been developed by G.~Narayanan 
mainly for the RSR data reduction. For each galaxy, the final spectrum is
constructed using the following procedure. First a set of scans of different pixels
were individually inspected in order to exclude data obtained under bad weather
conditions or with serious instrumental noise. Then continuum channels
are selected that exclude channels with {\sc CO} line emission and the
end channels affected by the bandpass roll-off. A linear baseline is
fitted over these continuum channels to calculate the rms noise of
each scan.
A set of scans for each object are summed weighted by the rms noise.
Lastly, a baseline is removed from the combined spectrum, using the 
same frequency range where we measure the rms of individual scans.
For most galaxies, a linear baseline has been fitted while a second order
fit has been adopted in some cases, especially for galaxies with low
signal-to-noise. The second order fit barely changes the rms over
the entire band ($\lesssim5\%$) but it helps to bring out weak
line features by modifying the local baseline structure around the
{\sc CO} line. The measured rms value of the fully processed 
spectrum for each galaxy is listed in Table~\ref{tbl-sample}.

In Figure~\ref{fig-rsr-a} (a)-(d), the coadded spectra of 29 ULIRGs
are shown, which have been divided into four groups depending on the
signal-to-noise. On the left-side of each object, the full band 
spectrum, the full band spectrum excluding noisy end channels is presented
in the same $T_A^*$ scale. Twenty seven objects in group
(a), (b), and (c) are qualified as detection by revealing a $>3\sigma$ feature 
within a few 100~km~s$^{-1}$ around the inferred {\sc CO} frequency from the 
optical redshift, which are marked with dotted lines in the full band 
spectra in Figure~\ref{fig-rsr-a} (a)-(d). Except for 2 ULIRGs in group (d)
which fail our definition of detection, the rest 27 have been classified
from (a) to (c) based on the following criteria:

\begin{enumerate}[(a)]
\item{{\it Bright} or {\it Intermediate}: In seven ULIRGs with 
S/N$>7$ (bright), and in another seven objects with 3.5$<$S/N$<7$,
$W_{\rm CO}>250~$km~s$^{-1}$ and the peak flux density $F^{\rm peak}_{\rm CO}>1.5~$ 
mK (intermediate), the lines can be picked out without any difficulty.}

\item{{\it Faint} or {\it Narrow}: Among eight ULIRGs with 3.5$<$S/N$<$5.5,
ﬁve objects with $F^{\rm peak}_{\rm CO}\lesssim1.5$ mK but $W_{\rm CO}>$200 
km~s$^{-1}$ (faint), and three objects with $W_{\rm CO}<$200 km~s$^{-1}$ yet
$F^{\rm peak}_{\rm CO}\gtrsim1.4$ mK (narrow) still show well deﬁned line features.}

\item{{\it Marginal}: There are five ULIRGs that still meet our detection
criteria but the lines are not as prominent as the other two groups
(1$<F^{\rm peak}_{\rm CO}<1.2$ mK). There could be some confusion with nearby
noise structure as indicated by the error in $W_{\rm CO}$ of $\gtrsim40$\%.}
\end{enumerate}

As seen in the full band spectra, the baselines are fairly flat 
throughout the entire frequency coverage with well behaved noise.
This is also demonstrated in Figure~\ref{fig-rms}
where the rms from the full band (Table~\ref{tbl-sample}) are compared
with the local rms which were measured using the continuum channels 
within $\sim1-2~$GHz range around the {\sc CO} line. In most cases, 
the rms are found along the equivalent line with $\lesssim10\%$ of variation.
 
It should be noted that such flat baselines over such wide bandwidths have
hitherto not been reported for millimeter wave receivers, and is a result
of the innovative design of the RSR.
The flat baseline of the RSR with well behaved noise will be 
particularly important for a blind line search of galaxies with no previous
redshift measurement, and hence this demonstration has been one of our
goals in this study. These characteristics of the RSR also allowed us to 
integrate down to detect fainter lines, yielding 93\% of the detection
rate including 5 marginal detections. The two non detections might be
due to the insufficient band coverage. In particular, \citet{gs99} reported
the {\sc CO} luminosity of IRAS~16487+5447 which is almost three times
larger than our upper limit based on the {\sc CO} line intensity upper
limit (3$\times$rms$\times$250~km~s$^{-1}$, \S~\ref{sec-measure}). However,
our baseline fit has successfully brought up the line close to the band
edge even in some cases with marginal signal-to-noise. Therefore we
suggest that those two galaxies may contain molecular gas of low flux 
density with a huge line width ($>500~$km~s$^{-1}$)
as found in some extreme cases like IRAS~F11180$+1623$ by \citet{dihn01}.

\section{Results}
\label{sec-res}
\subsection{Measuring the CO Quantities}
\label{sec-measure}
Integrated intensity ($I_{\rm CO}$), center frequency ($\nu_{\sc CO}$) and
{\sc CO} line width ($W_{\rm CO}$) are determined from the final 
reduced spectra. The integrated line intensity in K~km$s^{-1}$ is calculated
by summing the flux density of the line channels,
\begin{equation}
I_{\rm CO}=\sum_{line}F_{line}\frac{\Delta \nu}{\nu_{\sc line}} c \;\; K~ km~ s^{-1}
\end{equation}
where $F_{line}$ is flux density in K at each line channel, $\Delta \nu$ is the
channel width  (0.031 GHz for the RSR), $\nu_{line}$ is the central frequency of
each spectral channel in GHz, and $c$ is the speed of light in km~s$^{-1}$. 
The central frequency, $\nu_{\rm CO}$ has been calculated as a first moment of
the spectrum in the same frequency range used to measure $I_{\rm CO}$. The 
line width, $W_{\rm CO}$, is calculated as $I_{\rm CO}/F_{\rm peak}$. 
For the two non-detections, the quoted 3$\sigma$ upper-limits are
estimated assuming a line width of 250~km~s$^{-1}$ 
(cf. the mean line width $W_{\rm CO}$ of the 27 {\sc CO} detected
galaxies$\approx264\pm60$). 
The rms noise locally measured around the {\sc CO} line ($<2~$GHz, 
Table~\ref{tbl-obs}) has been adopted to calculate the upper limits. 
The measured {\sc CO} quantities are summarized in
Table~\ref{tbl-obs}.

%\placefigure{fig-rsr-c}
\addtocounter{figure}{-1}
\begin{figure*}
\includegraphics[bb=50 543 560 775]{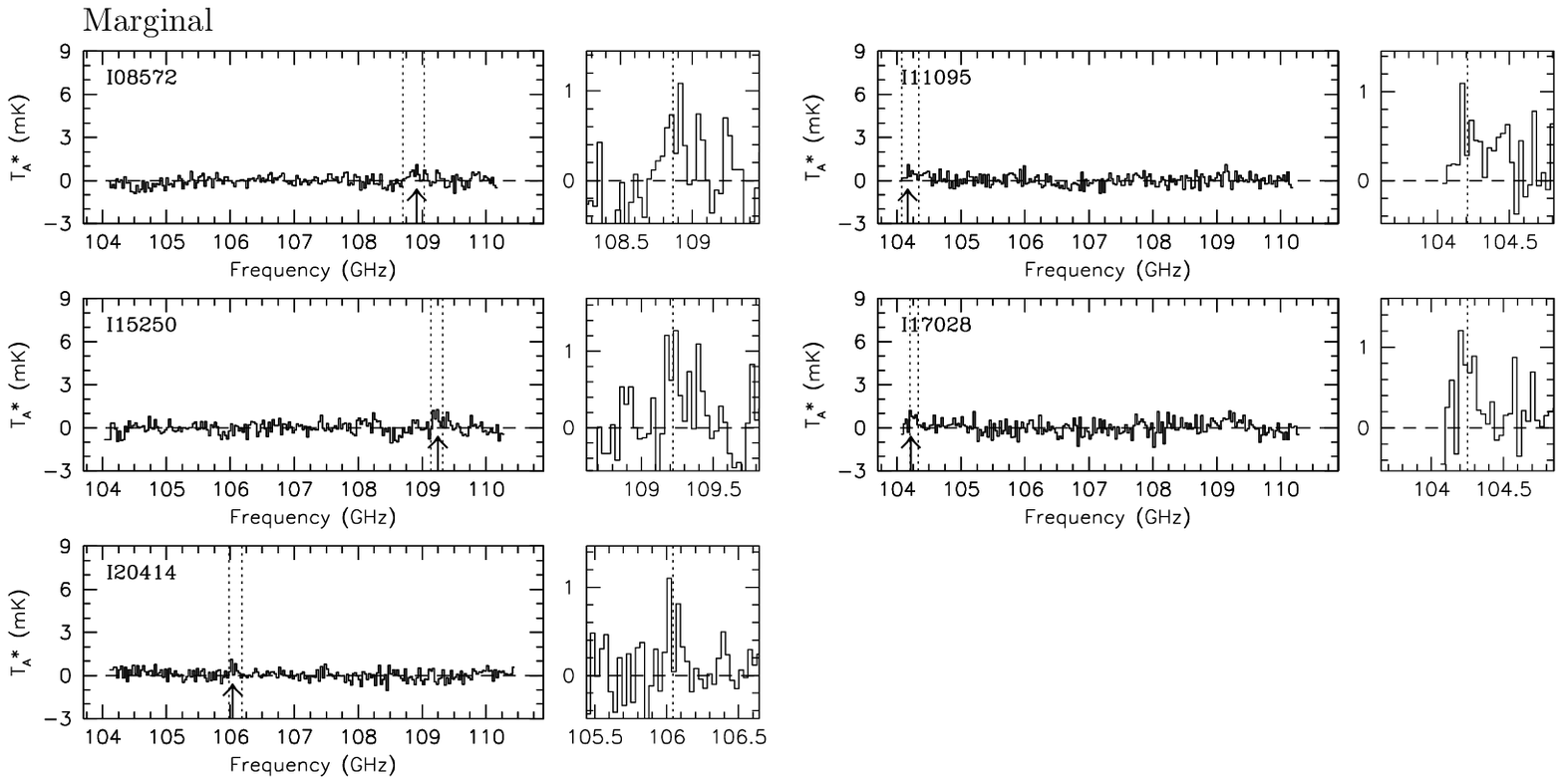}
\caption[]{(c) The RSR spectra of 5 ULIRGs of marginal 
detections. See the caption of Figure~\ref{fig-rsr-a}-(a) and
\S\ref{sec-obs}.2 for further details. \label{fig-rsr-c}}
\end{figure*}

%\placefigure{fig-rsr-d}
\addtocounter{figure}{-1}
\begin{figure*}
\includegraphics[bb=50 690 560 775]{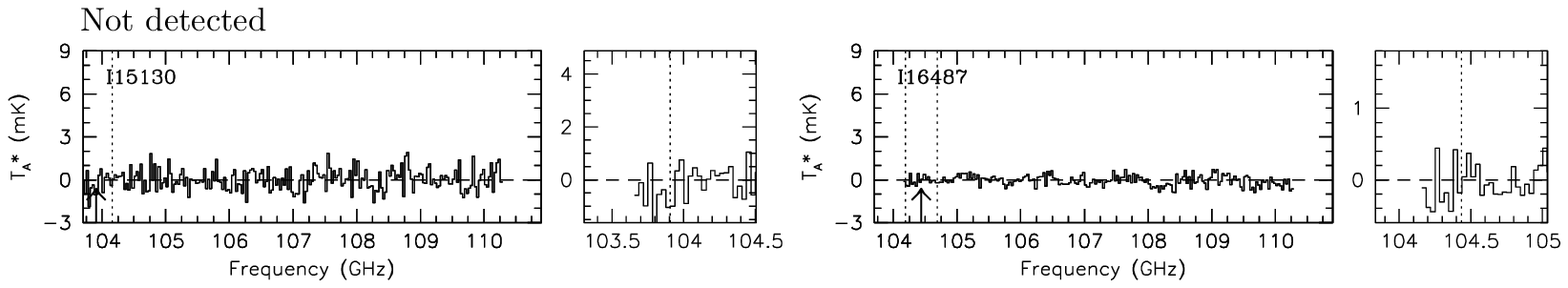}
\caption[]{(d) The RSR spectra of 2 ULIRGs which were not detected
in our study. The dotted lines in the full-band spectra indicate 
the 1~GHz width ($\sim2600~$km~s$^{-1}$) around the inferred {\sc CO}
frequency the optical velocity. The y-axis in the zoom-in spectrum
corresponds to $-3\sigma$ to $5\sigma$ as presented in
Table~\ref{tbl-sample}. 
\label{fig-rsr-d}}
\end{figure*}

%\placefigure{fig-rms}
\begin{figure}
\plotone{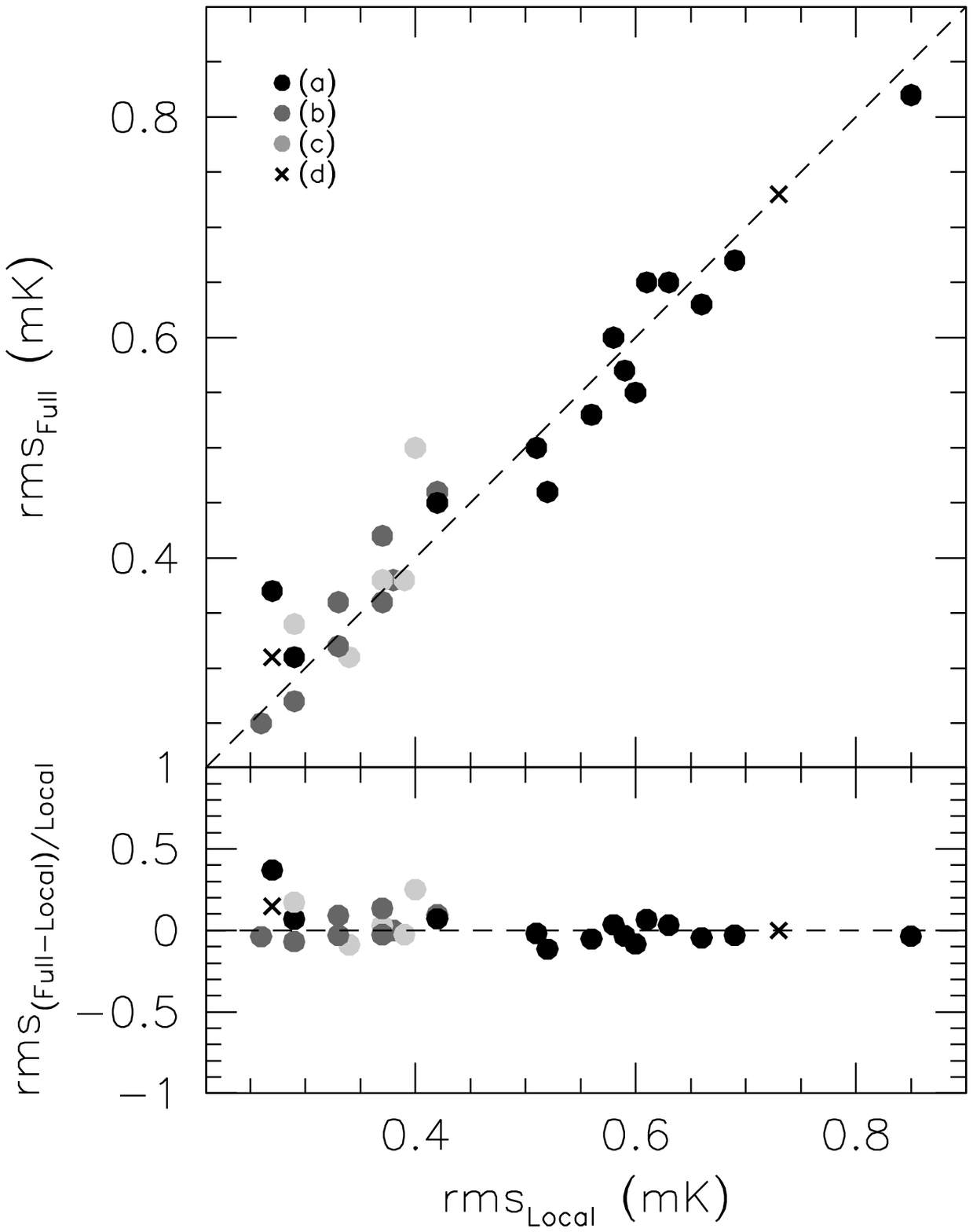}
\caption[]{Top) A comparison of the full band noise in the coadded 
RSR spectra to the local rms measured around the {\sc CO} line 
($<2$~GHz). Bottom) The percentage differnce in the full band rms and 
the local rms as a function of the local rms.
\label{fig-rms}}
\end{figure}

%\placefigure{fig-compub}
\begin{figure} 
\plotone{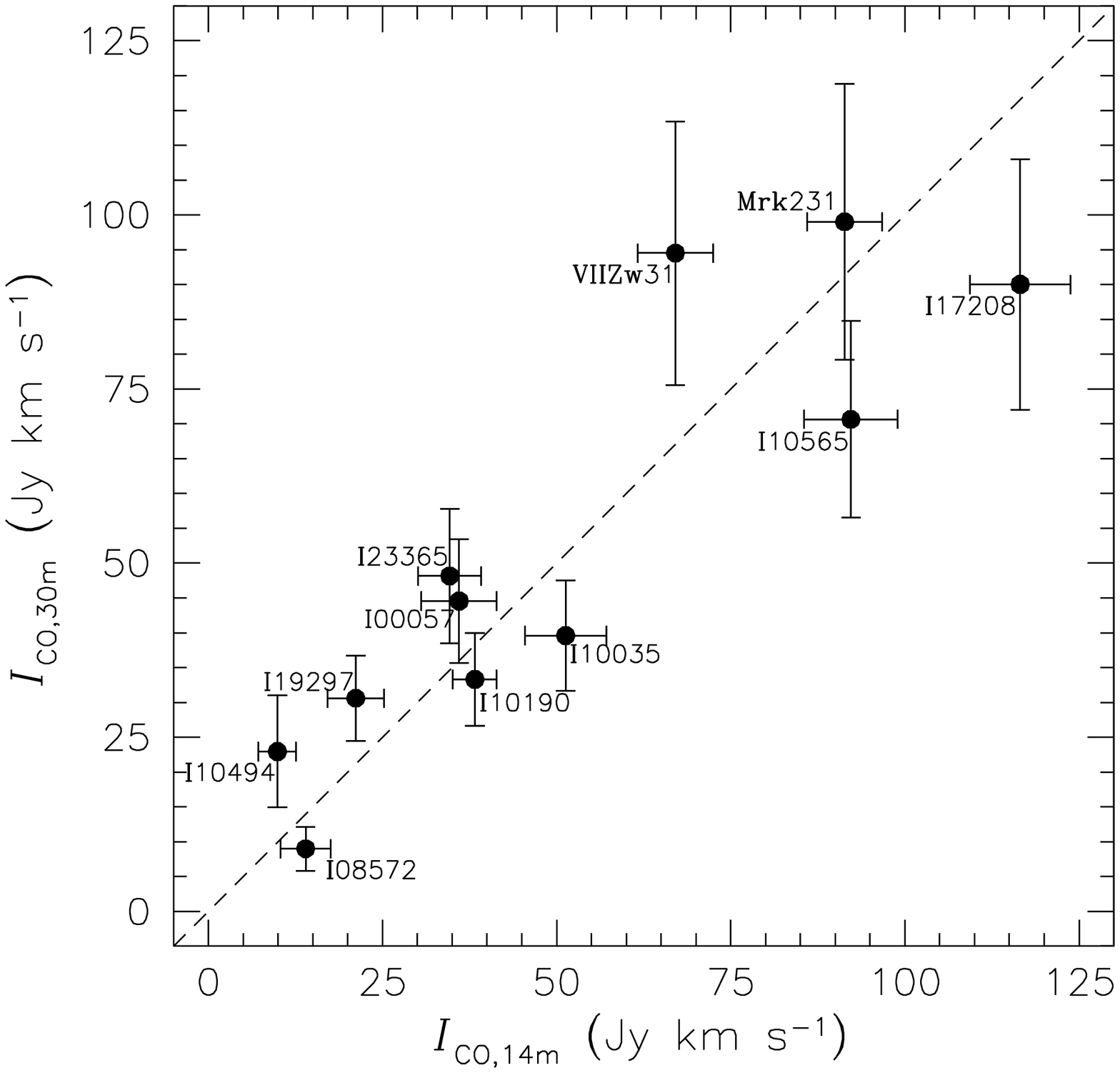}
\caption{Comparisons of $I_{\rm CO}$ from the FCRAO 14m telescope
with those measured by the IRAM~30m telescope \citep{solomon97}.
The antenna gain, G$=$45 and 4.5~Jy~K$^{-1}$ in $T_A^\prime$ and $T_{mb}$ 
have been adopted for the 14m and 30m telescope, respectively.
The 1-$\sigma$ error bars are shown for both measurements.
The dashed line shown is simply a linear relation 
that corresponds to $I_{\rm CO,30m}=I_{\rm CO,14m}$.
\label{fig-compub}}
\end{figure}

%\placetable{tbl-obs}
\begin{table}
\begin{center}
\caption{\sc $^{12}$CO~$J=1\rightarrow 0$ Line Parameters\label{tbl-obs}}
\begin{tabular}{lccccr}
\hline\hline
\multicolumn{1}{l}{Object} &
\multicolumn{1}{c}{rms} &
\multicolumn{1}{c}{$I_{\rm CO}$} &
\multicolumn{1}{c}{$\nu_{\rm CO}$} &
\multicolumn{2}{c}{$W_{\rm CO}$} \\
\multicolumn{1}{l}{} &
\multicolumn{1}{l}{mK} &
\multicolumn{1}{r}{K km s$^{-1}$} &
\multicolumn{1}{c}{GHz} &
\multicolumn{2}{c}{km s$^{-1}$} \\
\hline
I00057 &0.51& 0.80$\pm$0.12 & 110.36$\pm$0.02 & 347$\pm$ &\hspace{-0.2cm} 91\\ 
I05083 &0.61& 1.49$\pm$0.12 & 109.38$\pm$0.01 & 177$\pm$ &\hspace{-0.2cm} 20\\ 
I05189 &0.63& 0.66$\pm$0.16 & 110.56$\pm$0.04 & 258$\pm$ &\hspace{-0.2cm} 93\\ 
I08572 &0.34& 0.31$\pm$0.08 & 108.86$\pm$0.02 & 290$\pm$ &\hspace{-0.2cm}116\\
I09111 &0.56& 0.65$\pm$0.12 & 109.33$\pm$0.07 & 302$\pm$ &\hspace{-0.2cm} 94\\ 
I10035 &0.59& 1.14$\pm$0.13 & 108.30$\pm$0.01 & 320$\pm$ &\hspace{-0.2cm} 63\\ 
I10173 &0.38& 0.26$\pm$0.07 & 109.93$\pm$0.01 & 195$\pm$ &\hspace{-0.2cm} 81\\ 
I10190 &0.29& 0.85$\pm$0.07 & 107.13$\pm$0.04 & 378$\pm$ &\hspace{-0.2cm} 62\\
I10494 &0.26& 0.22$\pm$0.06 & 105.58$\pm$0.10 & 213$\pm$ &\hspace{-0.2cm} 77\\ 
I10565 &0.58& 2.05$\pm$0.15 & 110.54$\pm$0.00 & 291$\pm$ &\hspace{-0.2cm} 33\\ 
I11095 &0.37& 0.32$\pm$0.10 & 104.26$\pm$0.02 & 288$\pm$ &\hspace{-0.2cm}138\\ 
I12112 &0.66& 1.43$\pm$0.16 & 107.50$\pm$0.03 & 239$\pm$ &\hspace{-0.2cm} 37\\ 
I12540 &0.60& 2.03$\pm$0.12 & 110.60$\pm$0.02 & 243$\pm$ &\hspace{-0.2cm} 21\\ 
I13539 &0.29& 0.37$\pm$0.06 & 104.01$\pm$0.05 & 223$\pm$ &\hspace{-0.2cm} 53\\  
I14348 &0.85& 1.06$\pm$0.20 & 106.55$\pm$0.09 & 281$\pm$ &\hspace{-0.2cm} 82\\ 
I14394 &0.37& 0.45$\pm$0.10 & 104.38$\pm$0.07 & 325$\pm$ &\hspace{-0.2cm}113\\ 
I15130 &0.73&   $<$0.54     &~......          &~~~...... &\hspace{-0.2cm}   \\ 
I15250 &0.39& 0.30$\pm$0.08 & 109.27$\pm$0.01 & 238$\pm$ &\hspace{-0.2cm} 96\\ 
I15462 &0.42& 0.25$\pm$0.08 & 104.83$\pm$0.05 & 138$\pm$ &\hspace{-0.2cm} 59\\
I16487 &0.27&   $<0.21$     &~......          &~~~...... &\hspace{-0.2cm}   \\ 
I17028 &0.40& 0.34$\pm$0.10 & 104.30$\pm$0.04 & 281$\pm$ &\hspace{-0.2cm}145\\ 
I17132 &0.52& 1.07$\pm$0.10 & 109.67$\pm$0.04 & 378$\pm$ &\hspace{-0.2cm} 72\\ 
I17208 &0.69& 2.59$\pm$0.16 & 110.54$\pm$0.02 & 294$\pm$ &\hspace{-0.2cm} 28\\ 
I18470 &0.37& 0.32$\pm$0.09 & 106.86$\pm$0.01 & 243$\pm$ &\hspace{-0.2cm}105\\ 
I19297 &0.27& 0.47$\pm$0.09 & 106.20$\pm$0.11 & 308$\pm$ &\hspace{-0.2cm} 96\\ 
I20414 &0.29& 0.22$\pm$0.08 & 106.10$\pm$0.11 & 201$\pm$ &\hspace{-0.2cm} 96\\ 
I22491 &0.33& 0.26$\pm$0.06 & 107.03$\pm$0.02 & 184$\pm$ &\hspace{-0.2cm} 62\\ 
I23327 &0.33& 0.28$\pm$0.08 & 104.06$\pm$0.09 & 220$\pm$ &\hspace{-0.2cm} 94\\ 
I23365 &0.42& 0.77$\pm$0.10 & 108.30$\pm$0.03 & 251$\pm$ &\hspace{-0.2cm} 49\\ 
\hline
\end{tabular}
\end{center}
\end{table}

%\placefigure{fig-comspec}
\begin{figure*}
\plotone{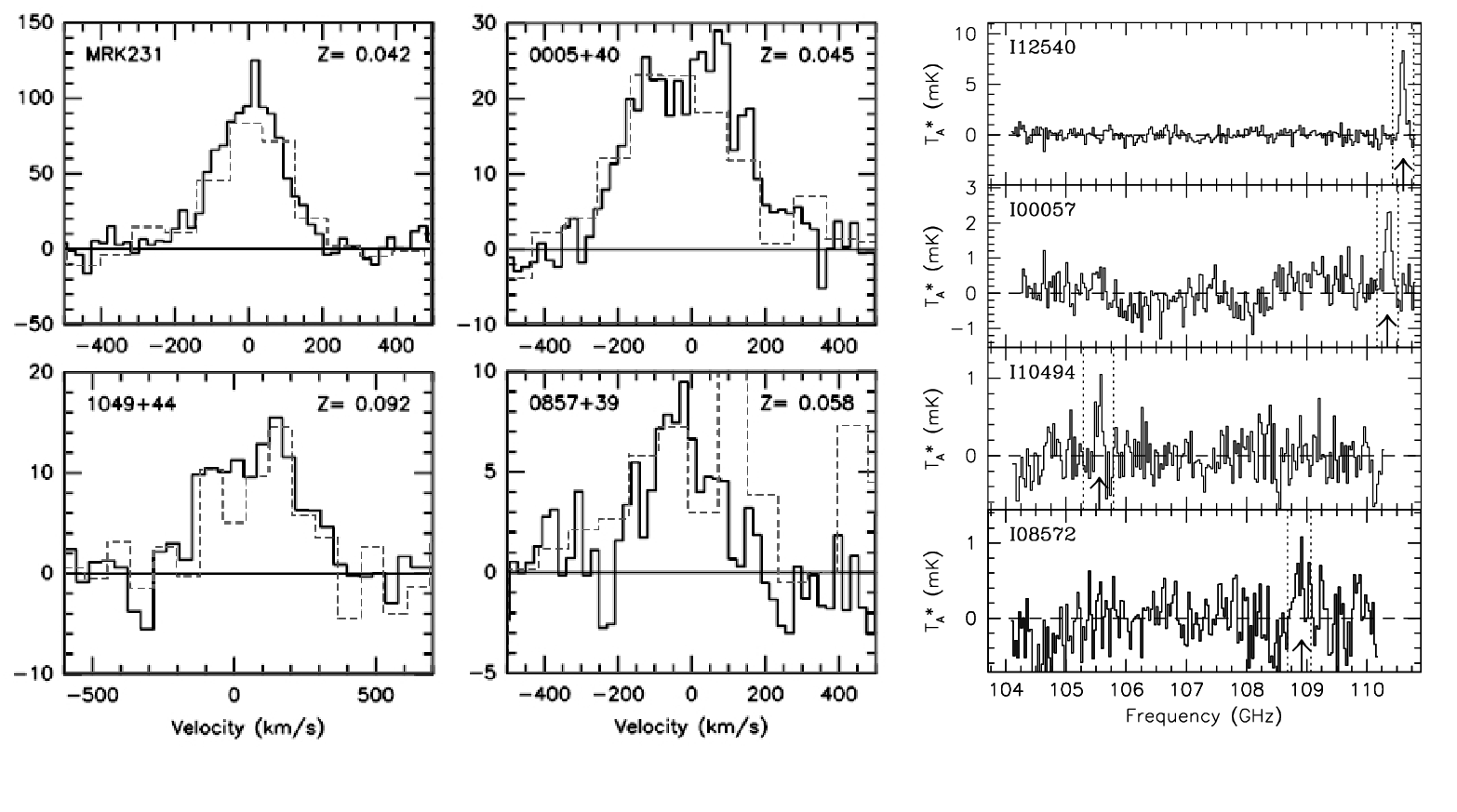}
\caption{Left) Comparisons of the 14m (dashed line) with 
30m spectra \citep[solid; ][]{solomon97}. In the upper row, one
example of {\sc CO} bright (I12540) ULIRG and one with intermediate
{\sc CO} line strength (I00057) are shown. In the bottom, each example
show a {\sc CO} faint object (I10494) and one marginal
detection (I08572). The spectra from IRAM~30m is shown in 
mK in $T_{mb}$ scale as presented in \citet{solomon97}; Right)
The RSR full-band spectra of the four ULIRGs are presented.
The y-axis (in mK in $T_A^*$ scale) has been scaled down with the
peak {\sc CO} flux density of each case. The frequency ranges blocked
with dotted lines are the regions compared with the study of 
\citet{solomon97} on the left-side. The upper arrow represents the 
optical redshift as Figure~\ref{fig-rsr-a}.\label{fig-comspec}}
\end{figure*}

%\placefigure{fig-histo}
\begin{figure}
\plotone{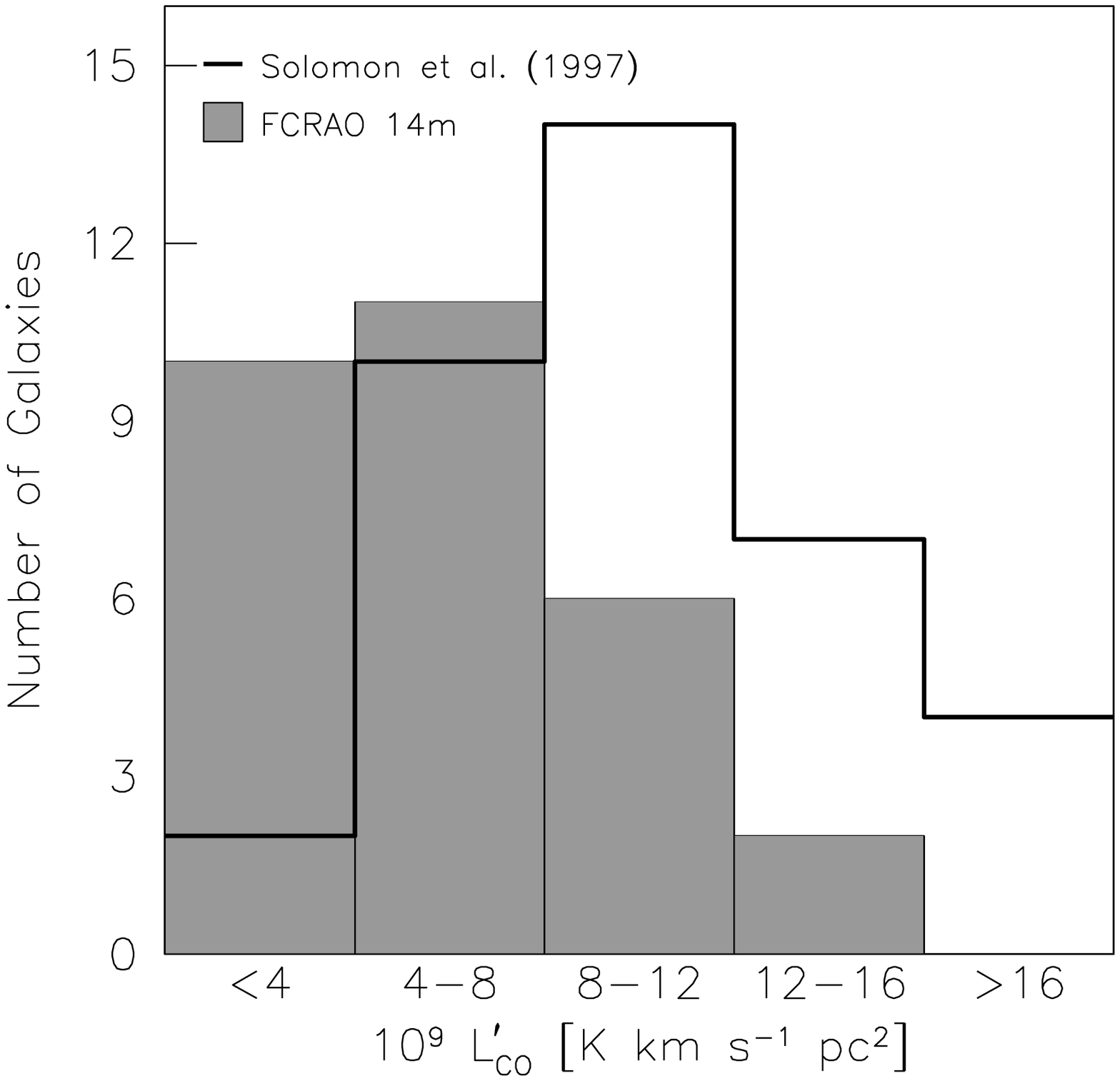}
\caption{Distributions of the sample with different ranges of {\sc
CO} luminosities. The thick line represents 37 low-z ULIRGs
($z<0.3$) from \citet{solomon97} and the shaded area represents 29
local ULIRGs ($z\lesssim0.1$) from this work. \label{fig-histo}}
\end{figure}

\subsection{Comparisons with Previous Measurements}
\label{sec-quality}
Among the prior {\sc CO} studies of LIRGs/ULIRGs, \citet{solomon97}
collected coherent {\sc CO} data using the IRAM~30m telescope of 37
ULIRGs and is the most comparable with our study. A comparison of 
our $I_{\rm CO}$ of 11 ULIRGs which were also observed by 
\citet{solomon97} is shown in Figure~\ref{fig-compub}.
Gain factors $G=$45 and 4.5 Jy~K$^{-1}$ in $T_A^*$ and $T_{mb}$ 
scale are adopted for the 14m and 30m telescope, respectively.  
In Table~\ref{tbl-cofir}, $I_{\rm CO}$ from the 14m telescope is given 
in Jy~km~s$^{-1}$. As shown in Figure~\ref{fig-compub}, most of
the objects are within their $1\sigma$ measurement uncertainty from 
the $I_{\rm CO,30m}=I_{\rm CO,14m}$ line and $I_{\rm CO}$ from the 
two telescopes are overall in good agreement. 

In Figure~\ref{fig-comspec}, we directly compare the 14m spectra 
with those from the 30m for four
objects with a range of signal-to-noise: one {\sc CO} bright object
(I12540, also known as Mrk~231), one with intermediate brightness 
(I00057), one {\sc CO} faint object (I10494), and one with a marginal 
detection (I08572). As shown on the left side of the figure,
the 14m spectra are generally in good agreement in spite of
the low spectral resolution ($\sim$90~km~s$^{-1}$ vs. 24 or
48~km~s$^{-1}$ for 30m). There may be some confusion in the cases of
marginal detections although the discrepancy in the total flux
is expected not more than 40\% inferred from the comparison of 
I08572. Lastly, the redshifts determined from our {\sc CO}
spectra (Table~\ref{tbl-cofir}) are consistent with optical redshifts for
the 27 detected sources, with $\Delta <V_{CO}-V_{opt}>=121\pm93~$km~s$^{-1}$.   

A comparison of the distribution of {\sc CO} luminosities for our survey
sample and those in the same redshift range measured by \citet{solomon97} 
is shown in Figure~\ref{fig-histo}.  This plot nicely shows that
nearly all of the ULIRGs in our sample are in the lower half
of the {\sc CO} luminosity bins.  Probing deeper into the lower {\sc CO} 
luminosity objects, we find a broader range of $L_{FIR}/L_{CO}$ ratio
for the ULIRG population, including the discovery of a handful of ULIRGs 
with extremely large $L_{FIR}/L_{CO}$ ratios ($>250~L_\odot/L_l$).  This newly identified
population is discussed further in \S~\ref{sec:COproperties}.

\subsection{CO Luminosity and Molecular Gas Mass \label{sec-lco}}

The {\sc CO} line luminosity, $L^\prime_{\rm CO}$,  is calculated as 
\begin{equation}
\label{eq-lco}
L^\prime_{\rm CO}=3.25\times10^7I_{\rm CO}\nu_{obs}^{-2}D_L^2~ (1+z)^{-3}~K~km~s^{-1}~pc^2
\end{equation}
where $I_{\rm CO}$ is the velocity integrated {\sc CO} intensity in Jy~km~s$^{-1}$, 
$D_L$ is the luminosity distance in Mpc, $\nu_{obs}$ is the observed frequency
in GHz, and $z$ is the redshift of the object \citep{solomon92}. We used the 
redshift determined using the {\sc CO} line to calculate $D_L$. For the undetected
sources, $L^\prime_{\sc CO}$ upper limits are computed using the upper limits
of $I_{\rm CO}$ in Table~\ref{tbl-obs} and their optical redshifts.
Derived {\sc CO} luminosities are listed in Table~\ref{tbl-cofir}. 

The {\sc CO} luminosity of our sample varies by more than a factor of 10, ranging 
1.2 -- 15.3$\times10^9~$K~km~s$^{-1}$~pc$^2$. Adopting the {\sc H$_2$}-to-{\sc CO} 
conversion factor for starburst systems 
\citep[$M_{gas}/L^\prime_{\rm CO}\approx0.8~M_\odot$~{[K~km~s$^{-1}$~pc$^2$]}$^{-1}$;][]{ds98}, 
$L_{\rm CO}^\prime$ of our sample corresponds to 1 -- 12$\times10^{9}~M_\odot$ of {\sc H$_2$},
which is comparable to normal to gas-rich spiral galaxies \citep{gs04b}. 
The median $L_{\rm CO}^\prime$ of our sample, $6--7\times10^{9}$~K~km~s$^{-1}$~pc$^2$, is about
three times as large as the $L^*$ value of the local {\sc CO} luminosity function \citep{keres03},
consistent with a simple model that these ULIRGs are mergers of two massive spiral galaxies.

%\placefigure{fig-lfirco}
\begin{figure*} 
\plotone{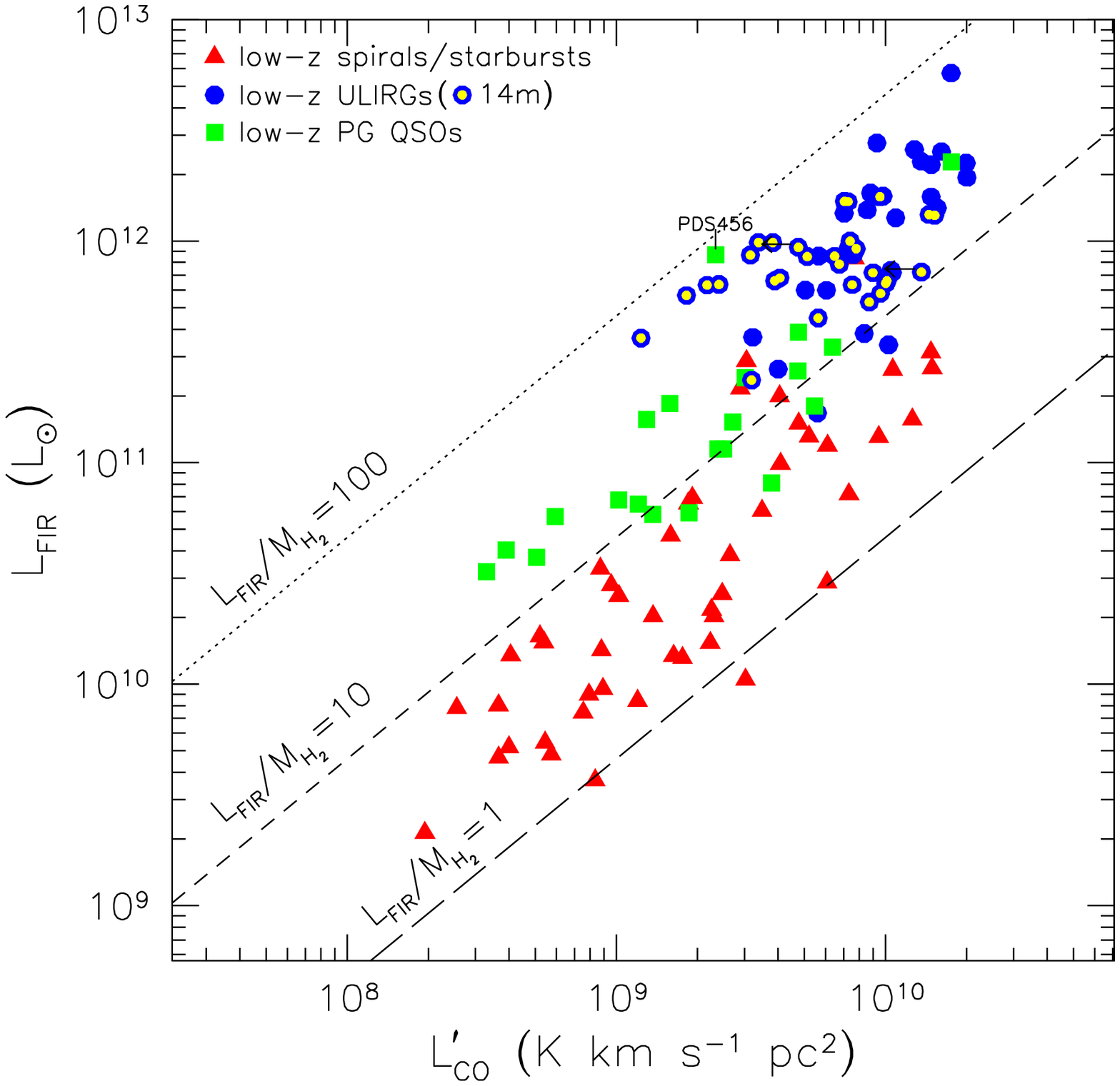}
\caption{$L_{FIR}$ vs. $L_{\rm CO}^\prime$ of low-$z$ galaxies with a range of far infrared 
luminosity. Spirals and starburst systems from \citet{gs04b} and
\citet{solomon97} are shown red triangles, $z\lesssim0.3$ Paloma- 
(PG) QSOs from \citet{alloin92}, 
\citet{evans01}, and \citet{scoville03} are shown by green squares.  ULIRGs by \citet{solomon97} 
are shown in blue circles. The yellow dots with a blue circle around represent the ULIRGs in our sample 
observed by RSR on the FCRAO 14m telescope. Non-detections are indicated with arrows.
Dotted and dashed lines indicate constant ratios of $L_{FIR}/M_{\rm H_2}=1$, 10, and 100 using 
the standard {\sc CO}-to-{\sc H$_2$} conversion ($M_{\rm H_2}$
[$M_\odot$]$=4.6L_{\rm CO}^\prime$ [K~km~s$^{-1}~$pc$^2$]). 
\label{fig-lfirco}} 
\end{figure*}

%\placetable{tbl-cofir}
\begin{table}
\begin{center}
\caption{\sc Infrared and CO Properties of the Sample\label{tbl-cofir}}
\begin{tabular}{rrlrcrrrc}
\hline\hline
\multicolumn{1}{c}{Object} &
\multicolumn{1}{c}{\hspace{-0.17cm}$I_{\rm CO}$} &
\multicolumn{1}{c}{$z^{\rm a}$} &
\multicolumn{1}{c}{$L_{\rm CO}^\prime$} &
\multicolumn{1}{c}{\hspace{-0.15cm}$\frac{f_{25}}{f_{60}}$} &
\multicolumn{1}{c}{$f_{60}$} &
\multicolumn{1}{c}{$f_{100}$} &
\multicolumn{1}{c}{$L_{FIR}$} &
\multicolumn{1}{c}{$q$}\\
\multicolumn{1}{c}{} &
\multicolumn{1}{c}{\hspace{-0.15cm}{\tiny (Jy~km/s)}} &
\multicolumn{1}{c}{} &
\multicolumn{1}{c}{\tiny ($10^{9}L_l$)$^{\rm b}$} &
\multicolumn{1}{c}{} &
\multicolumn{1}{c}{\tiny(Jy)} &
\multicolumn{1}{c}{\tiny(Jy)} &
\multicolumn{1}{c}{\tiny ($10^{12}L_\odot$)} &
\multicolumn{1}{c}{}\\
\hline
I00057&\hspace{-0.17cm}    36.0~~&0.044    &\hspace{-0.15cm}    3.18~~~&0.08&  4.47&  4.30& 0.24~~~& 2.853\\
I05083&\hspace{-0.17cm}    67.1~~&0.054    &\hspace{-0.15cm}    8.72~~~&0.10&  5.58&  9.62& 0.53~~~& 2.281\\
I05189&\hspace{-0.17cm}    29.7~~&0.043    &\hspace{-0.15cm}    2.41~~~&0.25& 13.70& 11.40& 0.65~~~& 2.729\\
I08572&\hspace{-0.17cm}    13.9~~&0.059    &\hspace{-0.15cm}    2.18~~~&0.23&  7.43&  4.59& 0.64~~~& 3.227\\
I09111&\hspace{-0.17cm}    29.3~~&0.055    &\hspace{-0.15cm}    3.88~~~&0.07&  7.08& 11.10& 0.68~~~& 2.829\\
I10035&\hspace{-0.17cm}    51.3~~&0.065    &\hspace{-0.15cm}    9.59~~~&0.06&  4.59&  6.24& 0.59~~~& 2.360\\
I10173&\hspace{-0.17cm}    11.7~~&0.049    &\hspace{-0.15cm}    1.24~~~&0.11&  5.80&  5.47& 0.37~~~& 2.800\\
I10190&\hspace{-0.17cm}    38.2~~&0.077    &\hspace{-0.15cm}   10.03~~~&0.11&  3.33&  5.57& 0.64~~~& 2.464\\
I10494&\hspace{-0.17cm}     9.9~~&0.092    &\hspace{-0.15cm}    3.82~~~&0.05&  3.53&  5.41& 0.99~~~& 2.375\\
I10565&\hspace{-0.17cm}    92.2~~&0.043    &\hspace{-0.15cm}    7.53~~~&0.09& 12.10& 15.10& 0.64~~~& 2.475\\
I11095&\hspace{-0.17cm}    14.4~~&0.106    &\hspace{-0.15cm}    7.40~~~&0.13&  3.25&  2.53& 1.01~~~& 2.260\\
I12112&\hspace{-0.17cm}    64.3~~&0.073    &\hspace{-0.15cm}   15.25~~~&0.06&  8.50&  9.98& 1.33~~~& 2.658\\
I12540&\hspace{-0.17cm}    91.3~~&0.042    &\hspace{-0.15cm}    7.27~~~&0.27& 31.99& 30.29& 1.49~~~& 2.255\\
I13539&\hspace{-0.17cm}    16.6~~&0.109    &\hspace{-0.15cm}    9.00~~~&0.07&  1.83&  2.73& 0.73~~~& 2.366\\
I14348&\hspace{-0.17cm}    47.7~~&0.082    &\hspace{-0.15cm}   14.53~~~&0.07&  6.87&  7.07& 1.33~~~& 2.372\\
I14394&\hspace{-0.17cm}    20.2~~&0.105    &\hspace{-0.15cm}   10.14~~~&0.18&  1.95&  2.39& 0.67~~~& 1.798\\
I15130&\hspace{-0.17cm} $<$24.3~~&0.109$^*$&\hspace{-0.15cm}$<$13.61~~~&0.20&  1.92&  2.30& 0.73~~~& 2.362\\
I15250&\hspace{-0.17cm}    13.5~~&0.055    &\hspace{-0.15cm}    1.83~~~&0.18&  7.29&  5.91& 0.57~~~& 2.809\\
I15462&\hspace{-0.17cm}    11.2~~&0.100    &\hspace{-0.15cm}    5.13~~~&0.16&  2.92&  3.00& 0.86~~~& 2.489\\
I16487&\hspace{-0.17cm}  $<$9.5~~&0.104$^*$&\hspace{-0.15cm} $<$4.75~~~&0.07&  2.88&  3.07& 0.93~~~& 2.254\\
I17028&\hspace{-0.17cm}    15.3~~&0.106    &\hspace{-0.15cm}    7.79~~~&0.04&  2.43&  3.91& 0.96~~~& 2.400\\
I17132&\hspace{-0.17cm}    48.2~~&0.051    &\hspace{-0.15cm}    5.63~~~&0.11&  5.68&  8.04& 0.45~~~& 2.446\\
I17208&\hspace{-0.17cm}   116.5~~&0.043    &\hspace{-0.15cm}    9.54~~~&0.05& 31.10& 34.90& 1.60~~~& 2.602\\
I18470&\hspace{-0.17cm}    14.4~~&0.079    &\hspace{-0.15cm}    4.06~~~&0.10&  4.07&  3.43& 0.69~~~& 2.566\\
I19297&\hspace{-0.17cm}    21.1~~&0.085    &\hspace{-0.15cm}    7.05~~~&0.08&  7.05&  7.72& 1.49~~~& 2.526\\
I20414&\hspace{-0.17cm}     9.9~~&0.087    &\hspace{-0.15cm}    3.38~~~&0.08&  4.36&  5.25& 0.98~~~& 2.346\\
I22491&\hspace{-0.17cm}    11.7~~&0.078    &\hspace{-0.15cm}    3.15~~~&0.10&  5.44&  4.45& 0.86~~~& 3.026\\
I23327&\hspace{-0.17cm}    12.6~~&0.108    &\hspace{-0.15cm}    6.74~~~&0.11&  2.10&  2.81& 0.79~~~& 2.405\\
I23365&\hspace{-0.17cm}    34.6~~&0.064    &\hspace{-0.15cm}    6.48~~~&0.11&  7.09&  8.36& 0.85~~~& 2.578\\
\hline		
\end{tabular}
\tablenotetext{a}{The redshift measured by using the {\sc CO} line. 
For non-{\sc CO} detections, an optically measured $z$ is quoted and indicated with asterisk.}
\tablenotetext{b}{$L_l\equiv$K~km~s$^{-1}$~pc$^2$}
\end{center}
\end{table}

\section{Discussion} 
\label{sec-dis} 

\subsection{Non-linear Correlation between $L_{FIR}$ and $L^\prime_{\rm CO}$ \label{sec:COproperties}} 

Using the fluxes in the $IRAS$ Faint Source Catalog, we have calculated  
$FIR$ luminosity, $L_{FIR}$  following \citet{lonsdale85},
\begin{equation} 
\label{eq-lfir}  
L_{FIR}=4\pi D^2_L~FIR \;\;L_\odot
\end{equation} 
$D_L$ is the luminosity distance (see \S~\ref{sec-lco}) and $FIR$ is the far-infrared flux,
\begin{equation} 
\label{eq-lfir}  
{FIR}=1.26\times10^{-14}~(2.58f_{60}+f_{100})~W~m^{-2},
\end{equation} 
where $f_{60}$ and $f_{100}$ are flux densities at 60 and 100 $\micron$ in Jy, respectively.
Considering the large uncertainty in dust emissivity, we have not applied any color
corrections to $L_{FIR}$ in Table~\ref{tbl-cofir}.

There is a well-known correlation between $IR$ luminosity and {\sc CO} luminosity for star
forming galaxies, presumably driven by the scaling relation between star formation
activity and the amount of fuel available \citep[see review by][]{ys91}.  This correlation
is reproduced in Figure~\ref{fig-lfirco}, including the 29 new {\sc CO} measurements of ULIRGs
obtained using the RSR system.  Galaxies with normal to high star formation rate
\citep{solomon97,gs04b} and Palomar-Green (PG) QSOs \citep{alloin92,evans01,scoville03} 
are also shown for comparison.
As previously reported \citep[e.g.][]{sanders91,solomon97,gs04b}, ULIRGs are the most luminous 
CO emitters in the local universe, with their {\sc CO} luminosity matching or exceeding those of the 
most luminous spirals.  Unlike most late type galaxies that follow a nearly linear 
$L_{FIR}$--$L^\prime_{\rm CO}$ correlation ($L_{FIR}/M_{\rm{H_2}}$=1-10~$L_\odot/M_\odot$) 
however, ULIRGs are over-luminous in the far-$IR$ for their {\sc CO} luminosity 
($L_{FIR}/M_{\rm{H_2}}$=10-100~$L_\odot/M_\odot$), 
leading to a significant up-turn in the $L_{FIR}$--$L^\prime_{\rm CO}$ correlation.

Our new survey of 29 ULIRGs has increased the number of ULIRGs with
{\sc CO} measurements to a total of 56. The resulting improvement in the total
sample size and statistics provides a better definition of the $L_{\rm CO}^\prime$
distribution as shown in Figure~\ref{fig-lfirco}.  By focusing on the CO-bright and
more luminous systems at slightly higher redshifts ($z\la0.3$), \citet{solomon97}
established the deviation of ULIRGs from the late type galaxies in the
field in the observed $L_{FIR}$--$L^\prime_{\rm CO}$ correlation.  
Our study has imporved the statistics particularly on the lower {\sc CO} luminosity 
end of the nearby sample at $z\la0.1$. 

PG QSOs as a group have $L_{FIR}/L^\prime_{\rm CO}$ = 67$\pm$35 $L_\odot/$[K~km~s$^{-1}$], 
intermediate between
ULIRGs (145$\pm$79) and late type field galaxies (22$\pm$21).  
As the result of probing the low {\sc CO} luminosity of the local ULIRG population, 
ULIRGs and PG QSOs
now appear to connect more smoothly in the $L_{FIR}$--$L^\prime_{\rm CO}$
correlation plot (Fig.~\ref{fig-lfirco}).  The well known correlation between
supermassive black holes (SMBHs) and their bulge stellar mass/velocity
dispersion \citep{tremaine02,marconi03} suggests that the formation/growth 
of SMBHs and their stellar hosts may be coupled.  Previously, some doubt was
expressed on the ULIRG-QSO connection owing to the apparent discrepancy in
their $L_{FIR}/L^\prime_{\rm CO}$ ratios \citep[e.g.,][]{yun04}.
However, with a better agreement revealed by our new ULIRG {\sc CO} survey,
this apparent discrepancy (driven by poor statistics and bias in the earlier
surveys) is no longer a serious challenge for the ULIRG-to-QSO
evolutionary scenario (see 5.2 for further discussion).

\begin{figure}
\plotone{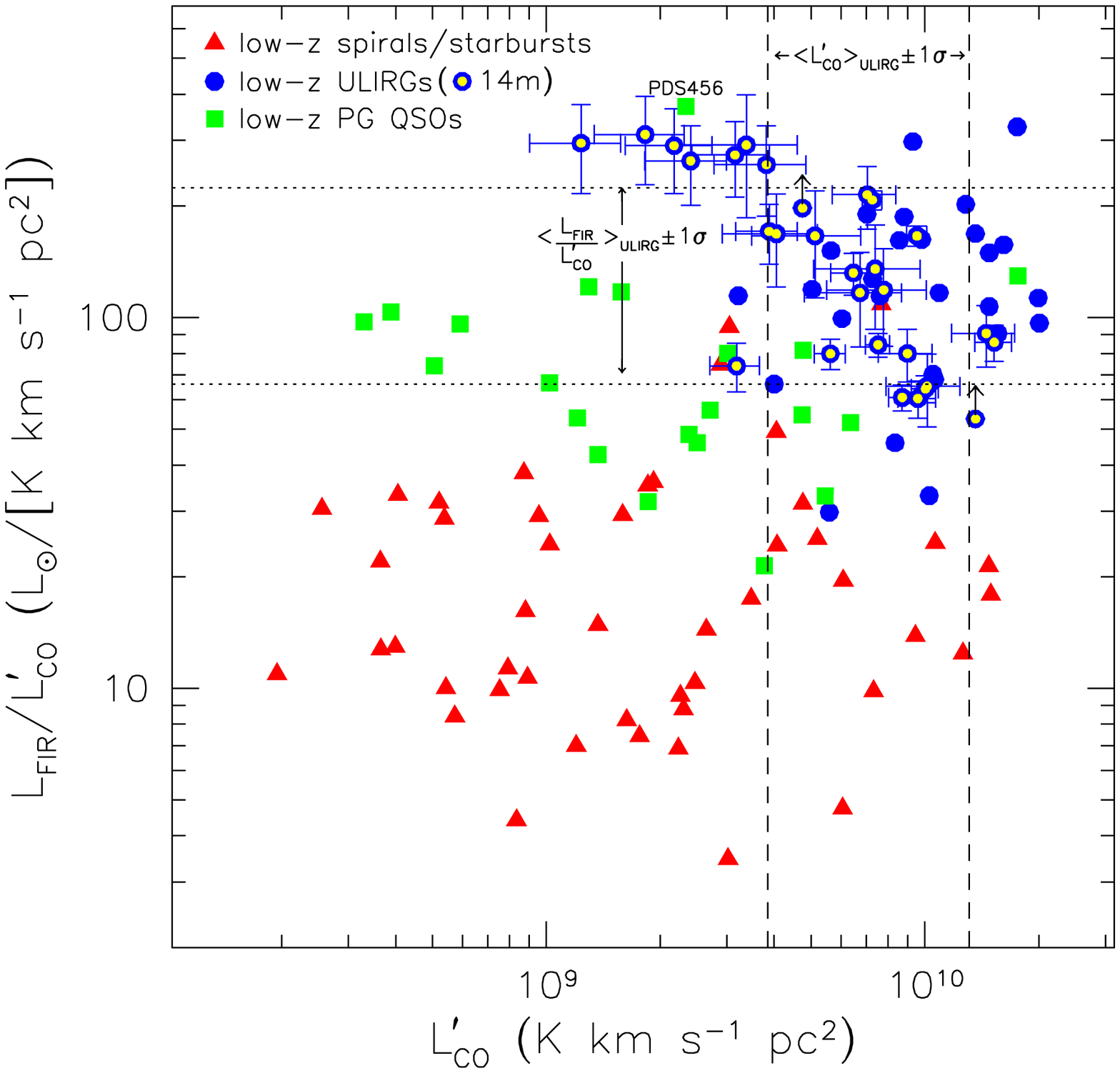}
\caption{$L_{FIR}/L^\prime_{\rm CO}$ vs. $L^\prime_{\rm CO}$. The same colored 
symbols are used as Figure~\ref{fig-lfirco}. The dashed lines and the 
dotted lines represent $L^\prime_{\rm CO}\pm1\sigma$ ($8.5\pm4.6\times10^9$ 
K~km~s$^{-1}$~pc$^2$) and $L_{FIR}/L^\prime_{\rm CO}$ (145$\pm$79 $L_\odot$ 
[K~km~s$^{-1}$~pc$^2$]$^{-1}$) of 56 ULIRGs collected in this study.
The error bars of our sample reflect the 1$\sigma$ uncertainties in our
{\sc CO} measurements. 
\label{fig-fcolco}}
\end{figure}

A small number of ULIRGs with extreme ratios of
$L_{FIR}/L_{\rm CO}^\prime\gtrsim 250$ $L_\odot/$[K~km~s$^{-1}$] have been identified by our new
ULIRG {\sc CO} survey (Fig~\ref{fig-lfirco} and \ref{fig-fcolco}).  
PDS~456, the most luminous QSO  in the local universe at 
$z=0.184$, and an ULIRG at $z=0.059$, I08572 were the only luminous infrared
sources previously identified with such an extreme $L_{FIR}/L^\prime_{\rm CO}$ ratio
\citep{yun04}. \citet{yun04} have speculated that these sources may be extremely
rare ULIRG-to-QSO transition objects based on the warm $IR$ color and small 
$L_{FIR}/L^\prime_{\rm CO}$ ratio.  Our study has identified
seven additional sources, and they are discussed further below, 
in \S~\ref{sec-evolution}.

%\placefigure{fig-fcolco}

%\placefigure{fig-firagn}
\begin{figure}
\plotone{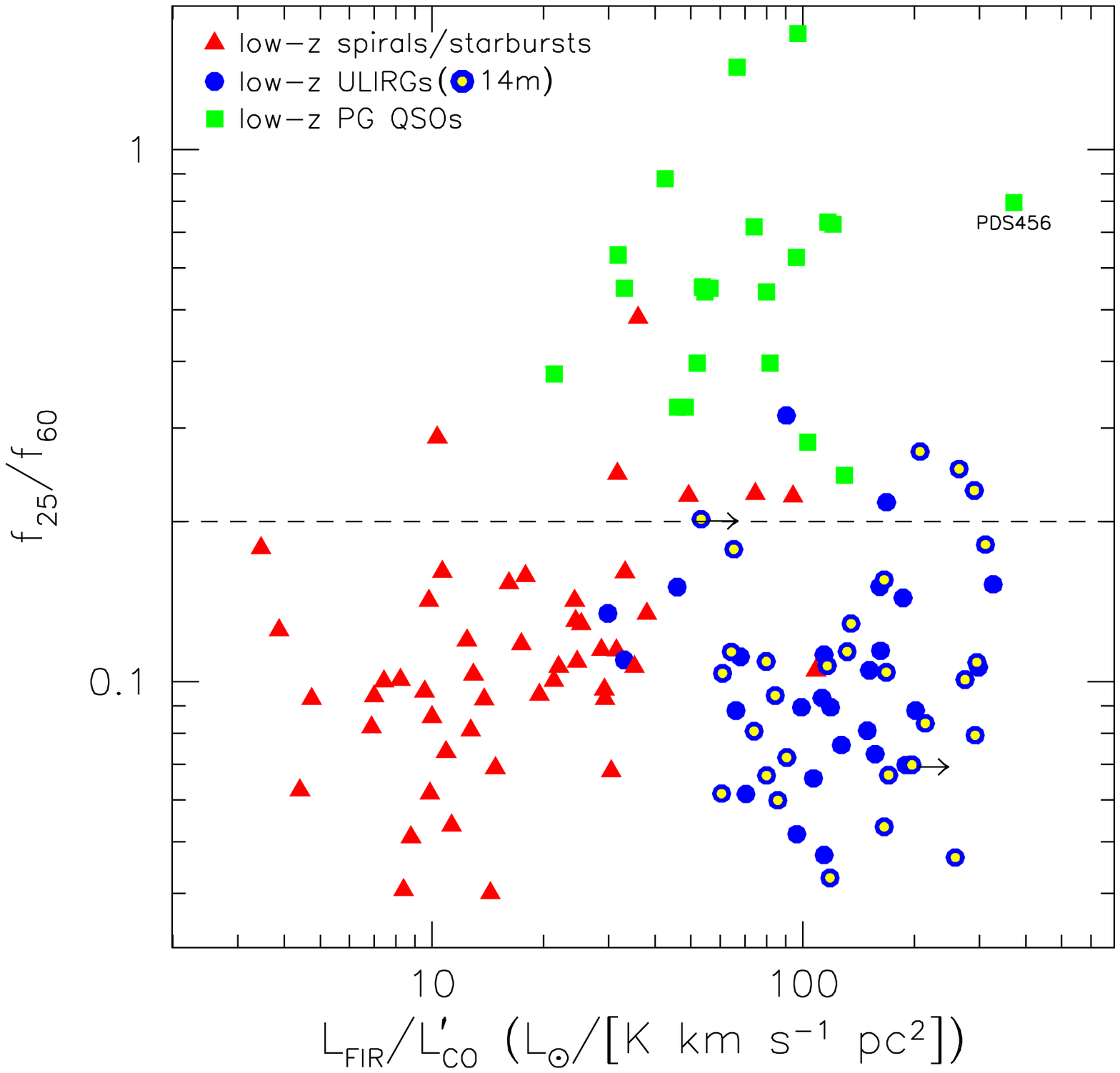}
\caption{Infrared color ($f_{25}/f_{60}$) vs. $L_{FIR}/L^\prime_{\rm CO}$. 
The same colored symbols are used as as Figure~\ref{fig-lfirco}. The dashed line represents
$f_{25}/f_{60}=0.2$, the borderline of warm and cool ULIRGs. 
\label{fig-firagn}}
\end{figure}

\subsection{AGN Activities as the Origin of Large $FIR$-to-{\sc CO} Luminosity Ratios for ULIRGs}

A number of $IR$ luminous galaxies display characteristics of Seyfert activity
in their optical spectrum, indicating that dust heating by AGN activities may
make an important contribution to the $IR$ emission. Although the actual frequency
is poorly determined \citep{veilleux06}, about 5\% of $IR$ bright
galaxies with $L_{IR}=10^{10-11} L_\odot$ and up to 50\% of the sources with
log $L_{IR}\geq 10^{12.3} L_\odot$ show broad, high excitation lines
characteristic of AGN activities \citep[][]{veilleux95,ks98,veilleux99,kewley01}.
An optical spectrum characteristic of Seyfert activity is found in 8 out of 29
galaxies in our sample and 12 out of 29 in the comparison sample of local ULIRGs
\citep{solomon97,gs04a}. Including those with a LINER spectrum (11 and 8),
respectively) would increase the fraction of ULIRGs with AGNs. 

Given that the bulk of ULIRG luminosity emerges as dust-processed IR emission,
optically thin tracers are required to probe the presence of AGN activities with
a greater accuracy.  Also, since all massive galaxies are expected to host a SMBH 
\citep[e.g.,][]{tremaine02,marconi03}, demonstrating that AGN activity
is the primary source of luminosity is far more important than establishing
the presence of a hidden AGN. For example, \citet{nagar03} have reported that
a high fraction ($\sim$75\%) of ULIRGs with Seyfert or LINER emission lines
host compact, non-thermal radio sources among the IRAS 1~Jy ULIRG sample.   
In contrast, the analysis of the Infrared Space Observatory (ISO) mid-$IR$ spectra 
\citep{genzel98,rigop99} and $IR$ photometry \citep[e.g.][]{farrah03} have 
shown that starbursts dominate the luminosity output in most ULIRGs while
only 20-30\% of ULIRGs are mainly powered by a central
AGN in the low-$z$ Universe.

Since the $L_{FIR}/L_{\rm CO}^\prime$ may be an indicator of the efficiency
of converting gas mass into luminosity, we explore whether this ratio may offer
a useful insight on the nature of the powering sources for the large $IR$
luminosity of ULIRGs. In Figure~\ref{fig-firagn}, we plot mid-$IR$ color
$f_{25}/f_{60}$, as a proxy of luminous obscured AGN \citep{degrijp85},
against the $L_{FIR}/L_{\rm CO}^\prime$ ratio for the local ULIRGs, PG QSOs,
and spirals. These three populations are fairly well separated in
this figure with only a small overlap among them.  Spirals and ULIRGs
have similar mid-$IR$ colors, but ULIRGs are distinguished by their larger
$L_{FIR}/L_{\rm CO}^\prime$ ratios.  PG QSOs show much warmer mid-$IR$ colors
($f_{25}/f_{60}\geq$0.2), and their $L_{FIR}/L_{\rm CO}^\prime$ ratios are
comparable to or somewhat smaller than those of ULIRGs.  Only $\sim$10\% of
the ULIRGs show mid-$IR$ colors as warm as QSOs.  The ULIRGs with Seyfert or 
LINER spectra are not clearly distinguishable by their mid-$IR$ color.
%little correlation is found between the $IR$ color and the presence of AGNs. 
ULIRGs as a group are clearly distinct from PG QSOs in both
$f_{25}/f_{60}$ color and in their $L_{FIR}/L_{\rm CO}^\prime$ ratios.
These trends imply that the SEDs of most ULIRGs are distinct from
those of QSOs and that their mid-$IR$ emission may be indicative of
dust heating by a starburst rather than AGN activities.  Compared
with non-ULIRG starburst systems, however, ULIRGs show a higher efficiency 
of converting molecular gas into stars as indicated by their high 
$L_{FIR}/L_{\rm CO}^\prime$ as other studies have previously found. 

\begin{figure}
\plotone{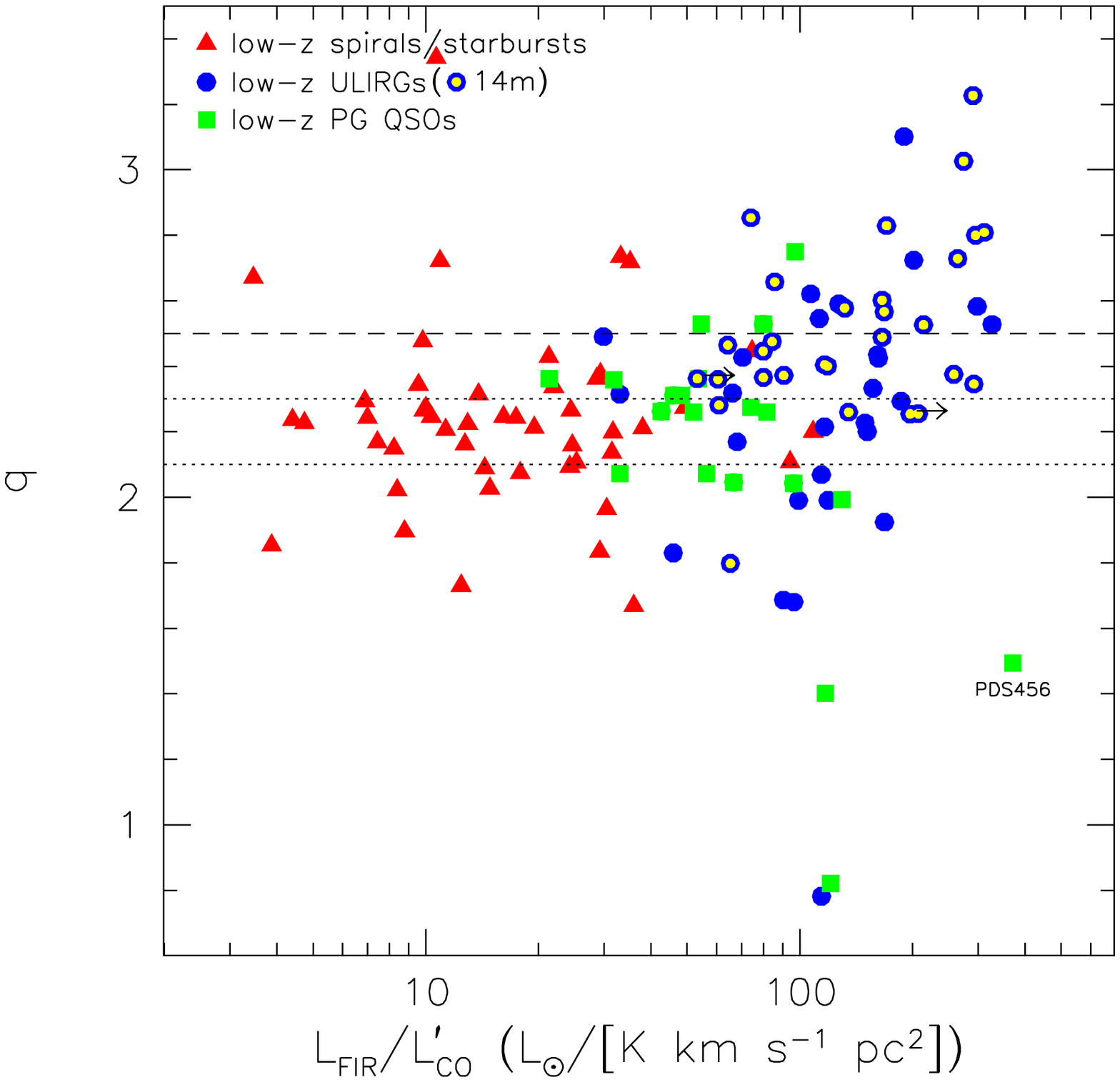}
\caption{A plot of radio-to-FIR luminosity ratio (``$q$-factor''). The same
colored symbols are adopted as Figure~\ref{fig-lfirco}. The long-dashed and
dotted lines represent the mean $q$ and the scatter of \citet{condon92}'s sample
of spiral galaxies (2.3$\pm$0.2). 
\label{fig-qplot}}
\end{figure}

Another commonly used optically thin indicator of AGN activity is a compact, 
non-thermal nuclear radio source that produces radio emission in excess of 
the expected trend from the well-known radio-$IR$ correlation 
\citep[``radio-excess'' with $q\ll 2.1$; see][]{condon92,yun01}.  
For example, the majority of the high redshift QSOs detected both in mm/submm
(rest frame far-$IR$) show radio continuum well in excess of their far-$IR$
luminosity \citep{yun00,petric03,wang07}. A quantitative measure of the
logarithmic FIR-to-radio flux density ratio, which is known as
{\lq\lq}$q${\rq\rq}-parameter, is derived using the following definition,
\begin{equation}
q\equiv log\frac{FIR}{3.75\times10^{12}~{\rm W~m^{-2}}}
       -log\frac{S_{1.4}}{{\rm W~m^{-2}~Hz^{-1}}}
\end{equation}
where $S_{1.4}$ is 1.4~GHz radio flux density in W~m$^{-2}$~Hz$^{-1}$
\citep[see ][]{condon92}. Lower $q$ values correspond to higher AGN activity.
In Figure~\ref{fig-qplot} we plot $q$-parameter against $L_{FIR}/L_{\rm CO}^\prime$.
As mentioned earlier, \citet{nagar03} have reported a high frequency
($\sim$75\%) of compact, non-thermal radio sources among the IRAS 1~Jy
ULIRG sample with Seyfert or LINER emission lines.  Among the 29 ULIRGs
we studied, however, only one object (I14394) shows a clear radio-excess,
and the radio-excess ULIRGs are rare among the local ULIRG population in
general (see Fig.~\ref{fig-qplot} and Table~\ref{tbl-AGN}).  In the local
Universe, only about 1\% of the $IR$-selected galaxies
show clear radio-excess \citep{yun01}, and only about 10\% of optically
selected QSOs are radio-loud \citep{kellermann89}.  Therefore, we can conclude
that there is little evidence for a high level of radio AGN activity among
the ULIRG population or alternatively can infer a low duty cycle of the radio 
AGN phase.  

%\placefigure{fig-qplot}

%\placefigure{fig-firpro}
\begin{figure}
\plotone{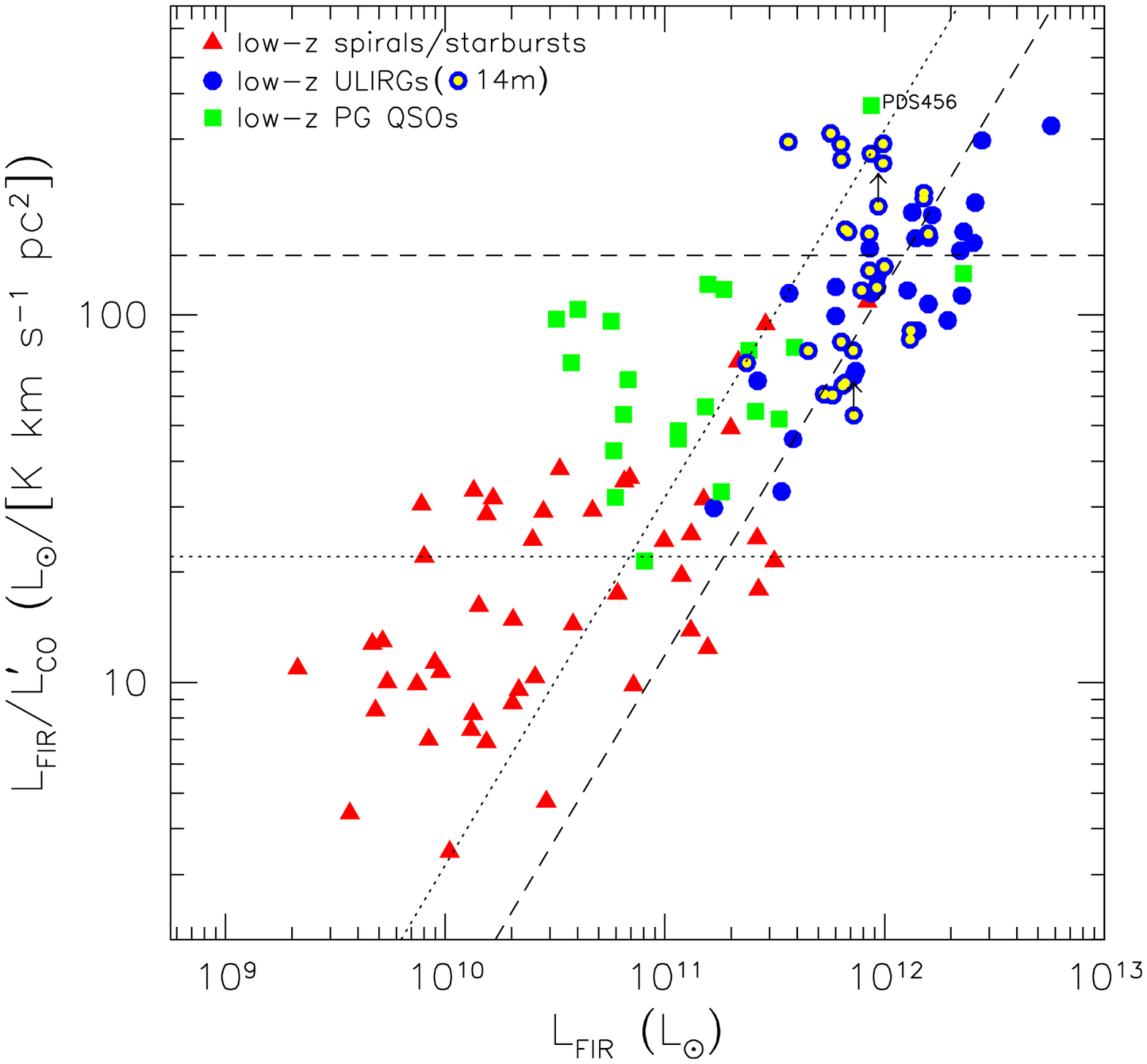}
\caption{$L_{FIR}/L^\prime_{\rm CO}$ vs. $FIR$ luminosity. 
The dashed lines represent the typical star formation effeiciency (SFE) of 
the ULIRG sample (145 $L_\odot/L_l$ $\approx$ 116 $L_\odot/M_\odot$) and 
$L^\prime_{\rm CO}$ ($8.47\times10^9$~$L_l$), respectively. Those of the 
spiral sample (22 $L_\odot/L_l$ $\approx$ 4--5 $L_\odot/M_\odot$, and $3.15\times10^9$ $L_l$)
are indicated with dotted lines (the conversion factors for different populations
are given in \S~\ref{sec-dis}.4).
\label{fig-firpro}}
\end{figure}

\subsection{Merger-induced Starburst Activities as the Origin of Large $FIR$-to-{\sc CO} Luminosity Ratios for ULIRGs}

Since there is little direct evidence linking a large $FIR$-to-{\sc CO} 
luminosity ratio and AGN activity, we next explore whether the 
merger-driven starburst phenomenon can account for the large 
luminosity ratios observed. In Figure~\ref{fig-firpro}, we plot the
$L_{FIR}/L_{\rm CO}^\prime$ ratios of ULIRGs and non-ULIRG systems as
a function of $L_{FIR}$. A typical star formation rate of nearby
{\sc CO}-rich spirals (4~$L_\odot M_\odot^{-1}$) is similar to that
of galactic clouds and is generally attributed to  recent formation
of OB stars in a quiescent disk \citep[e.g.][]{ss88}. 
The $L_{FIR}/L_{\rm CO}^\prime$ ratio of 20~$L_\odot M_\odot^{-1}$, 
shown by a dotted line, is more typical of local starburst galaxies.
\citet{yun04} and others have noted that PG QSOs display a similar
$L_{FIR}/L_{\rm CO}^\prime$ ratio and have proposed a closer physical and 
evolutionary link with the local starbursts, rather than with the ULIRGs.  
Most local ULIRGs have a $L_{FIR}/L_{\rm CO}^\prime$ ratio  well above
the value of non-ULIRG starburst systems, implying that the ULIRG
phenomenon not only requires the largest amount of molecular gas found
in the most gas-rich systems in the local universe, but it also
requires a much more efficient mechanism that converts the gas mass to
luminosity. Another potentially important clue to the underlying
physical mechanism is that the dispersion in the $FIR$-to-{\sc CO}
luminosity ratio for the ULIRGs is the largest among the different
populations compared. This dispersion may reflect varying evolutionary
stages of the ULIRG phase.

A useful insight on the observed $L_{FIR}/L_{\rm CO}^\prime$ ratio 
\citep[also dubbed ``star formation efficiency''][]{ys91} for the starbursts
and ULIRGs is offered by numerical studies of merging disk galaxies. 
Motivated by a high frequency of morphological peculiarities
associated with ULIRGs \citep{sanders88a,farrah01}, \citet{mh96} have
investigated the response of interstellar medium (ISM) during a merger
of two gas-rich spirals and the subsequent secular evolution of the
triggered starburst activity.  A particularly relevant result from
this study is that the compression and concentration of gas (details
of which are shown to be highly dependent on the initial conditions)
during the course of the merger lead to spikes of elevated starburst
activity lasting 5-20 $\times 10^6$ yr (see their Fig.~5).  
In a more recent simulational study, \citet{matteo08} have investigated
the enhancement in star formation during galaxy interactions,
suggesting that major mergers can increase the star formation 
efficiency (SFE) by a factor of a few tens compare to that
in unperturbed disks which appears to be consistent with
what is shown in Figure~\ref{fig-firpro}.

We interpret this prediction using our data by drawing lines of constant
{\sc CO} luminosity $L_{\rm CO}^\prime$ (long-dashed lines) in
Figure~\ref{fig-firpro}.  The long dashed line on the right side
corresponds to the average {\sc CO} luminosity of all ULIRGs, and
it goes through the upper envelope of the {\sc CO} luminosity associated 
with CO-rich spirals in the field. While the long dashed line on the left
corresponds to a $L_{\rm CO}^\prime$ 5 times smaller, approximately the 
$L^*$ value of the local {\sc CO} luminosity function
\citep[$10^9$~K~km~s${-1}$~pc$^2$;][]{keres03}. 

As one of the CO-rich field galaxies experience a large spike in the
massive star formation rate (induced by a galaxy-galaxy collision in
the cases of the Mihos \& Hernquist study), both $L_{FIR}$ and
$L_{FIR}/L_{\rm CO}^\prime$ ratio would increase along the right
long-dashed line, reaching the area occupied by the ULIRGs.  As the
starburst activity fades, the galaxy would climb back down along
the same line.  Therefore, the apparent broad trend between
$L_{FIR}/L_{\rm CO}^\prime$ ratio and $L_{FIR}$ seen in
Figure~\ref{fig-firpro} can be naturally explained by secular changes
in the starburst luminosity (externally triggered or internally by
dynamical instabilities such as bars or spiral density waves) to the
gas mass ratio.

One may expect the evolutionary trajectory to be somewhat skewed from
the right long-dashed line since the starburst activity would consume
some of the gas and would possibly remove additional gas through
energetic feedback processes \citep[e.g.,][]{heckman90,martin05}.
On the other hand, the resulting higher gas pressure will lead to an
increase in {\sc CO} excitation and line luminosity, partly
compensating for the gas mass loss --- see the discussions of elevated
{\sc CO} luminosity in nuclear starburst regions by \citet{solomon97}
and \citet{ds98}.  These competing effects should contribute to the
tightness in the observed correlation.

%\placetable{tbl-AGN}
\begin{table}
\begin{center}
\caption{\sc AGN Diagnostics for the Extreme $L_{FIR}/L'_{CO}$ ULIRGs\label{tbl-AGN}}
\begin{tabular}{lcccc}
\hline\hline
\multicolumn{1}{l}{Object} &
\multicolumn{1}{c}{$L_{FIR}/L^\prime_{\rm CO}$} &
\multicolumn{1}{c}{$f_{25}/f_{60}$} &
\multicolumn{1}{c}{$f_{1.4}$} &
\multicolumn{1}{c}{$q$} \\
\multicolumn{1}{l}{} &
\multicolumn{1}{l}{} &
\multicolumn{1}{r}{} &
\multicolumn{1}{c}{(mJy)} &
\multicolumn{1}{c}{} \\
\hline
I05189 & 264 & 0.25 & 29 & 2.73 \\ 
I08572 & 290 & 0.23 & 4.8 & 3.22 \\
I10173 & 295 & 0.11 & 11 & 2.80 \\ 
I10494 & 258 & 0.05 & 20 & 2.38\\
I15250 & 311 & 0.18 & 13 & 2.81 \\ 
I20414 & 292 & 0.08 & 25 & 2.35\\
I22491 & 274 & 0.10 & 5.8 & 3.03\\
\hline
I03158 & 298 & 0.10 & 13 & 2.60 \\ 
I14070 & 327 & 0.07 & 5.5 & 2.54 \\ 
\hline
\end{tabular}
\tablenotetext{cf.}{The mean ratio of $L_{FIR}/L^\prime_{\rm CO}$ of
the ULIRG, the QSO, and the spiral sample collected in this study are
145 (113 excluding 9 extremers in the table), 67, and 22 $L_\odot/L_l$,
respectively.}
\end{center}
\end{table}

\subsection{Extreme $L_{FIR}/L_{\rm CO}^\prime$ Objects and ULIRG Evolution \label{sec-evolution}}

An important outcome of our ULIRG {\sc CO} study is the discovery
of a special group of ULIRGs; I10494, I05189, I22491, I08572, I20414, I10173, and
I15250 (in the order of increasing $L_{\rm CO}^\prime$). These systems 
show extreme $L_{FIR}/L_{\rm CO}^\prime$ ratios which ranges from 250 to 310
$L_\odot/L_l$ (Fig.~\ref{fig-lfirco}), which is greater than the mean of
the rest ULIRGs by a factor of two or more. With the standard conversion factor
for spiral galaxies, this corresponds to a star formation efficiency of 
$>50$ $L_\odot/M_\odot$ as shown in Figure~\ref{fig-lfirco}, and potentially 
$>300$ $L_\odot/M_\odot$, adopting the conversion factor for starburst
systems \citep{ds98}. All but one (I15250) had been previously observed 
in {\sc CO} by others (Table~\ref{tbl-sample}) and five were detected
but I20414 although they have not been recognized as a distinct group
before. The only other object previously known for such a large
$L_{FIR}/L_{\rm CO}^\prime$ ratio is PDS~456, the most luminous QSO in
the local universe \citep[$z<0.3$][]{yun04}. These objects are readily 
identifiable in Figures~\ref{fig-lfirco}, \ref{fig-fcolco}, and
\ref{fig-firpro} by their extreme $L_{FIR}/L_{\rm CO}^\prime$ ratios, 
apart from the rest of the ULIRGs. Along with PDS~456, these seven ULIRGs 
have some of the lowest
far-$IR$ luminosity, suggesting that a comparatively larger fraction of
their total $IR$ luminosity arises in the mid-$IR$ wavelengths for a
given $L_{FIR}/L_{\rm CO}^\prime$. While the extreme
$L_{FIR}/L_{\rm CO}^\prime$ of these systems is primarily attributed
to low $L_{\rm CO}^\prime$, there are also two objects with a similar 
$L_{FIR}/L_{\rm CO}^\prime$ but higher $L_{\rm CO}^\prime$ among the sample
of \citet{solomon97}, I03158 and 14070.  These are two of the highest 
luminosity ULIRGs known, with nearly 10 times larger far-$IR$ luminosity
than the seven extreme ULIRGs identified in this study such that these may
be a different class of objects altogether. However, this luminosity
distinction may simply arise from the redshift restriction imposed by the 
RSR spectrometer.  Our sample includes the lowest redshift ULIRGS
corresponding to lower luminosities.

As a first step toward understanding the mechanism(s) responsible for 
their extreme $L_{FIR}/L_{\rm CO}^\prime$ ratio, we examine in detail 
the known properties of the individual sources, including the two from the
\citet{solomon97} sample: 

\begin{itemize}
\item{\it IRAS~03158$+$4227.} This $z=0.134$ ULIRG shows a second core
which is much fainter than the one in the center. While the optical
morphology is suggestive of a merging event, this object is found near
AGN dominated systems in \citet{spoon07}'s diagnostic diagram for
mid-infrared spectra of $IR$ galaxies, with a relatively small
equivalent width of the 6.2~\micron ~PAH emission feature 
and a weak 9.7 \micron ~silicate strength.

\item{\it IRAS~14070$+$0525.} This compact $z=0.264$ ULIRG does not
show any obvious signatures of interactions or merging. This system however,
has disturbed isophotes \citep{vks02} and morphologically classified as a
remnant by \citet{dasyra06b}. Its spectrum shows characteristics of Sy2
\citep{veilleux99}. 

\item{\it IRAS~05189$-$2524.} The optical image of this $z=0.043$ ULIRG
suggests that it is in a late stage of a merger with a highly obscured,
compact single nucleus \citep{farrah03}.  It is one of the six ULIRGs in
the IRAS Bright Galaxy Sample with the warmest $f_{25}/f_{60}$ color
\citep{sanders88a}. Presence of both an obscured AGN \citep{young96} and
a buried starburst \citep{dudley99} is suggested by a Sy2 spectrum
\citep{veilleux95} and an X-ray spectrum characteristic of a Compton
thin AGN and a thermal component \citep{risaliti00,lutz04}. 
By modeling the observed SED, \citet{farrah03} conclude that 
$>30$\% of the total $IR$ luminosity is contributed by the AGN.

\item{\it IRAS~08572$+$3915.} This $z=0.059$ ULIRG shows two cores
which are well separated by 6 kpc. Both the stellar nuclei
are dominated by H$\alpha$ emission \citep{cam05}. It is also a 
well-known warm object ($f_{25}/f_{60}=0.23$) with a strong
evidence for an AGN \citep{soifer00,farrah03}. 

\item{\it IRAS~10173$+$0828.} This $z=0.049$ ULIRG looks disky compared
to most ULIRGs with a highly asymmetric outer disk. It is a {\sc OH}
megamaser galaxy ({\sc OHM}), which is thought to be caused by
infrared radiation from the surrounding environment \citep{baan85}.
Most {\sc OHM}s are known to be warm (U)LIRGs \citep{chen07} although
this system is not a particularly warm ULIRG among the sample with
$f_{25}/f_{60}\approx0.11$.

\item{\it IRAS~10494$+$4424.} This $z=0.092$ ULIRG presents two
spikes to the north-northwest which \citep{vks02} suggest tidal origin.
In their SPITZER spectroscopic study, \citet{ima07} find no signature of 
a buried AGN in this system, which is consistent with the SED study by
\citet{farrah03}. The upper limit of the AGN contribution to the
total $IR$ flux estimated by \citet{farrah03} is less than $\sim$13\%.

\item{\it IRAS~15250$+$3609.} This $z=0.055$ ULIRG shows a ring-like
structure around a spheroid in the center \citep{scoville00} with a
much dimmer nucleus 0.7 arcsec away from the center \citep{farrah03}.
\citet{lutz99} have classified it as a starburst/LINER based on its
mid-$IR$ spectrum. However, in \citet{farrah03}'s SED study, this
object has been found with one of the largest AGN contribution
($L_{\rm AGN}/L_{IR}\gtrsim0.6$) among the sample.

\item{\it IRAS~20414$-$1651.} This $z=0.087$ ULIRG was observed in 
{\sc CO} by \citet{mirabel90} but not detected. It shows a main nucleus
with highly elongated isophotes \citep{vks02}. \citet{murphy96} find
a double core in the optical which appears to be one system in the
infrared. \citet{farrah03} find its AGN contribution to the total
$IR$ luminosity of $\sim13\%$, below the mean of their sample
(23\%). However, \citet{dasyra06b} find a black hole mass of
$\gtrsim10^8$ $M_\odot$ (larger than that of I05189 by a factor of
$\approx3.5$) based on the bulge dispersion relation \citep{tremaine02}.

\item{\it IRAS~22491$-$1808.} This $z=0.078$ ULIRG is a close (2.2 kpc) 
pair with two tidal tails \citep{vks02}. \citet{cui01} have proposed
a multiple merger origin for this system, which is also supported by
its mid-$IR$ spectrum \citep{lutz99}. However, this ULIRG is one of the
few objects with the largest AGN contribution ($L_{\rm AGN}/L_{IR}\gtrsim0.7$) 
in the sample of \citet{farrah03}'s SED study along with I15250.

\end{itemize}
Based on their high $L_{FIR}/L_{\rm CO}^\prime$ and evidence for AGNs
(except I03158 which has not been well studied, and I10494 with the lowest
$L_{FIR}/L^\prime_{\rm CO}$), we speculate that these ULIRGs
are currently consuming enormous amounts of molecular gas to feed the
central black holes, and are likely to become QSOs. The rarity of PDS~456-like
QSO however, suggests that the timescale of forming classical quasars 
in this way is relatively short compared to the timescale of growth of
an AGN in a ULIRG phase.

The interconnection between ULIRGs and QSOs has been suggested by
\citet[][b]{sanders88a} who have demonstrated the similarity between 
the SEDs of warm ULIRGs ($f_{25}/f_{60}>0.2$) and those of QSOs. Warm ULIRGs
also morphologically resemble QSOs with a more prominent spheroid, weaker
tidal features, and brighter nuclei compared to their cooler counterparts 
\citep{veilleux06}. These observations suggest that cool, starbursting 
ULIRGs may go through an AGN-like warm ULIRG phase and eventually
become optically selected QSOs once the 
burst of star formation decays and the nucleus sheds its obscuring dust
\citep{sanders88a,yun04,veilleux06,dasyra06a}. 

The transition from a ULIRG phase to a QSO phase can be driven by
rapid conversion of gas into stars and/or a subsequent growth of a 
supermassive black hole \citep{yun04}. Due to rapid consumption of
gas, transition objects are expected to deviate from the standard ULIRG
$L_{FIR}$-$L_{\rm CO}^\prime$ correlation in the way that results in a
larger $L_{FIR}$ for a given molecular gas mass compared to most
ULIRGs. There are only a few such ULIRGs that have been found in
previous {\sc CO} studies which are biased to luminous ULIRGs and 
one QSO, PDS~456 \citep{yun04}. In our study however, we find four 
additional ULIRGs with PDS~456-like $L_{FIR}/L_{\rm CO}^\prime$ 
along with I08572 which already has been known \citep{gs99}.

\section{Summary}
\label{sec-sum}
We have presented $^{12}${\sc CO}~$J=1\rightarrow 0$ observations of 
29 ULIRGs at $z$=0.043-0.11, using the Redshift Search Receiver (RSR). 
In total 27 systems have been detected including 9 new detections,
which has increased the number of local ULIRGs with {\sc CO}
measurement by 15\%. The {\sc CO} line luminosity
$L_{\rm CO}^\prime$ of our sample ranges from 1.2 to 
$15.3\times 10^9$~K~km~s$^{-1}$~pc$^2$. Adopting the {\sc CO}-to-{H$_2$}
conversion factor for ULIRGs/QSOs  
($M_{\rm H_2}$=0.8~$L^\prime_{\rm CO}$~$M_\odot$~[K~km~s$^{-1}$~pc$^2$]$^{-1}$),
the inferred cold gas mass in H$_2$ form in those ULIRGs is 
1 -- 12$\times 10^9~M_\odot$. We have investigated $L_{\rm CO}^\prime$
of local ULIRGs as function of $IR$ and radio properties.
\begin{enumerate}
\item{ULIRGs are 10-100 times higher in $L_{FIR}/L_{\rm CO}^\prime$ than
spirals/starburst galaxies and form a continuous track with low-$z$ QSOs,
consistent with results from previous studies. However as a result
of probing deeper in {\sc CO} faint objects, our survey finds a
broader range of $L_{FIR}/L_{\rm CO}^\prime$, resulting in a smoother
transition between the ULIRG and QSO populations. }
\item{ULIRGs are  well separated in $f_{25}/f_{60}$ vs. 
$L_{FIR}/L_{\rm CO}^\prime$ space from spirals/starburst systems by much
larger $L_{FIR}/L_{\rm CO}^\prime$. They are also distinct from QSOs by
their cooler mid-$IR$ color ($f_{25}/f_{60}<0.2$).}
\item{The radio excess, $q$ of ULIRGs is not significantly different 
from those of non-ULIRG populations.}
\item{Seven ULIRGS with extreme values of 
$L_{FIR}/L_{\rm CO}^\prime \gtrsim 250$ are identified.  These are 
similar to the luminous quasar PDS~456.}
\end{enumerate}
From these, we conclude that the power sources of most local ULIRGs are 
mainly sporadic starbursts which are likely to be driven by merging events,
rather than AGNs. However, we always find some evidence for a powerful 
central source in those ULIRGs with extreme $L_{FIR}/L_{\rm CO}^\prime$,
which may be representative of transition objects to QSOs. These objects
are not distinct from most ULIRGs in other properties which might be due
to a short timescale of the transition. 

\acknowledgments
We would like to thank the anonymous referee for a detailed review of the 
manuscript and useful comments that helped to improve the paper.
We are grateful to Mike Brewer, Don Lydon, Kamal Souccar, Gary Wallace,
Ron Grosslein, John Wielgus, Vern Fath, and Ronna Erickson for their technical
support of Redshift Search Receiver commissioning. This work was supported by
NSF grants AST~0096854, AST~0540852, and AST~0704966.

{}

\end{document}